\documentclass[fleqn,usenatbib]{mnras}

\usepackage[pdftex]{graphicx}

\usepackage{newtxtext,newtxmath}

\usepackage[T1]{fontenc}
\usepackage{ae,aecompl}

\usepackage{graphicx}	
\usepackage{amsmath}	

\title[Prospects of newly detecting SFGs]{Prospects of newly detecting nearby star-forming galaxies by the Cherenkov Telescope Array}

\author[Shimono et al.]{
Naoya Shimono,$^{1}$\thanks{E-mail: shimono@astron.s.u-tokyo.ac.jp}
Tomonori Totani,$^{1,2}$
Takahiro Sudoh$^{1}$
\\
$^{1}$Department of Astronomy, the University of Tokyo, 7-3-1 Hongo, Tokyo 113-0033, Japan\\
$^{2}$Research Center for the Early Universe, the University of Tokyo, 7-3-1 Hongo, Tokyo 113-0033, Japan\\
}

\date{Accepted XXX. Received YYY; in original form ZZZ}

\pubyear{2020}

\begin{document}
\label{firstpage}
\pagerange{\pageref{firstpage}--\pageref{lastpage}}
\maketitle

\begin{abstract}
Prospects of the Cherenkov Telescope Array (CTA) for the study of very
high energy gamma-ray emission from nearby star-forming galaxies are
investigated.  In the previous work, we constructed a model to
calculate luminosity and energy spectrum of pion-decay gamma-ray
emission produced by cosmic-ray interaction with the interstellar medium
(ISM), from four physical quantities of galaxies [star formation rate
 (SFR), gas mass, stellar mass, and effective radius]. The model is
in good agreement with the observed GeV--TeV emission of several nearby
galaxies. Applying this model to nearby galaxies that are not yet
detected in TeV (mainly from the KINGFISH catalog), their
hadronic gamma-ray luminosities and spectra are predicted.  We identify
galaxies of the highest chance of detection by CTA, including NGC
5236, M33, NGC 6946, and IC 342. Concerning gamma-ray spectra, NGC
1482 is particularly interesting because our model predicts that this
galaxy is close to the calorimetric limit and its gamma-ray spectral
index in GeV--TeV is close to that of cosmic-ray protons injected into
ISM. Therefore this galaxy may be detectable by CTA even though its GeV
flux is below the {\it Fermi} Large Area Telescope sensitivity limit. In the TeV regime,
most galaxies are not in the calorimetric limit, and the predicted TeV
flux is lower than that assuming a simple relation between the TeV
luminosity and SFR of M82 and NGC 253, typically by a factor of
15. This means that a more sophisticated model beyond the
calorimetric limit assumption is necessary to study TeV emission from
star-forming galaxies.
\end{abstract}




\section{Introduction}
\label{section:introduction}
All galaxies with substantial star formation activity (star-forming
galaxies, SFGs) are gamma-ray emitters. Gamma rays are produced by
the decay of pions, which are produced by collisions of cosmic-ray protons
accelerated by supernova remnants with the interstellar medium (ISM).
This process is dominant in the diffuse Galactic gamma-ray background
radiation \citep{Abdo2009}.  Observing this gamma-ray emission from
extragalactic galaxies is not easy because of smaller luminosities
compared with those of active galactic nuclei (AGNs) such as blazars,
but in recent years an increasing number of nearby SFGs [the Large and
  Small Magellanic Clouds (LMC and SMC), M31, NGC 253, M82, NGC 4945,
  NGC 1068, NGC 2146, Arp 220, Arp 299, and M33] have been detected in
the GeV band by the {\it Fermi} Large Area Telescope (LAT)
\citep{Abdo2010a,Abdo2010b,Abdo2010c,Abdo2010d,Ackermann2012,Tang2014,Peng2016,Lopez2018,Ajello2020,Xi2020b,Xi2020},
proving that gamma-ray emission from SFGs is indeed ubiquitous. M82
and NGC 253 were detected also in the TeV band by ground-based air
Cherenkov telescopes
\citep{Acero2009,VERITASCollaboration2009,H.E.S.S.Collaboration2018},
and more detections in TeV are expected by the Cherenkov Telescope
Array (CTA) project \citep{CherenkovTelescopeArrayConsortium2019}.  A large sample of gamma-ray emitting SFGs in
GeV--TeV in the near future will be useful to study the production and
propagation of cosmic rays in various types of galaxies
and to study high-energy background radiation. It is then
important to prepare theoretical predictions for gamma-ray luminosities
and spectra of nearby SFGs.

The most popular approach to predict gamma-ray luminosity from a
galaxy is to relate it with some of physical quantities of the galaxy,
especially the star formation rate (SFR). Gas mass of a galaxy is often
additionally used to take into account the interaction of cosmic rays with ISM
\citep{Torres2004,DomingoSantamaria2005,Persic2008,Lacki2010,Inoue2011,Lacki2011,Wang2018,Ambrosone2021}.
In other modelings,
microphysical parameters about cosmic-ray propagation (e.g. magnetic
field strengths, convection velocities, or mean free paths) are
introduced as free parameters and these are fit to reproduce the
observed gamma-ray luminosity of a galaxy
\citep{YoastHull2013,YoastHull2014,Eichmann2016,Peretti2019,Krumholz2020}.
In a more detailed approach, cosmic-ray production and
propagation are numerically simulated within a galaxy,
though it is difficult to adopt such an approach
to predict gamma-ray emission from a variety of nearby galaxies
\citep{Martin2014,Pfrommer2017,Chan2019}.

In a previous work, we constructed a model to predict gamma-ray
luminosity and spectrum of a SFG, from four input
parameters (SFR, gas mass, stellar mass, and effective radius) that
are easily available for nearby galaxies [\citet{Sudoh2018}, hereafter
S18]. The escape time scale of cosmic-ray protons is estimated as a
function of the proton energy, considering diffusion in turbulent
magnetic fields and advection. It was found that this model reproduces
the observed gamma-ray luminosities of nearby galaxies better than
those using only SFR and/or gas mass of galaxies.
In this paper, we apply the S18 model to nearby galaxies that
are not yet detected in the TeV band (mainly from the
KINGFISH catalog), predict their gamma-ray emission
properties, and discuss their relation to galactic physical
parameters. Then 
the prospect of future studies
with CTA will be discussed.

In Section~\ref{section:model} our model is described, 
especially about some minor 
changes from the original S18 model.
The nearby galaxy sample studied here will be presented in 
Section~\ref{section:sample}.  Then results are presented in Section~\ref{section:result},
followed by conclusions 
in Section~\ref{section:conclusions}.

\section{Model}
\label{section:model}
We use the model constructed by S18.
In this model, the proton injection spectrum is formulated as: 
\begin{equation}
\frac{dN}{dtdE_p}=C\left(\frac{\psi}{\mathrm{M_\odot\medspace yr^{-1}}}\right)\left(\frac{E_p}{\mathrm{GeV}}\right)^{-\Gamma_\mathrm{inj}}
\end{equation}
where $\psi$ is star formation rate, $\Gamma_{\rm inj}$ is proton
injection index, and $C$ is the parameter determined by fitting the
model to the observed GeV luminosities. 
We set the maximum energy of proton energy as ${ 4\times10^{15}\,{\rm eV}}$,
which corresponds to the knee.
This choice does not affect our results, because its effect starts to appear at the energy range where intergalactic absorption by extragalactic background light (EBL) is important (see below).
Then by a one-zone modeling
assuming disk geometry, we consider the probability of cosmic-ray proton
collisions with ISM as a balance between collision timescale and
escape timescale: $f=1-\exp(-t_{{\rm esc}}/t_{{\rm pp}})$.
Collision timescale, $t_{\rm pp}$, is expressed as $t_{\rm pp}=(n_{\rm gas}\sigma_{\rm pp} c)^{-1}$, 
where $n_{\rm gas}$ is ISM gas density and $\sigma_{\rm pp}$ is the pp cross-section \citep{Kelner2006}.
The escape time scale of cosmic-ray
protons is calculated as the minimum of advection timescale and
diffusion timescale: $t_{\rm esc}=\min(t_{\rm diff}, t_{\rm adv})$.
The advection timescale is estimated from the
disc scale height and velocity dispersion assuming gravitational
equilibrium: $t_{\rm adv}=H_{\rm g}/\sigma$ and $G\Sigma =\sigma^2/(2\pi H_{\rm g})$,
where $H_{\rm g}$ is scale height 
[estimated from the assumption that it is proportional to effective radius and MW values 
$R_{\rm eff, MW}=6.0\, {\rm kpc}$ \citep{Sofue2009} and $H_{\rm g, MW}=150\,{\rm pc}$ \citep{Mo2010}], 
$\sigma$ is the escape velocity 
and $\Sigma=(M_{\rm star}+M_{\rm gas})/(\pi R_{\rm eff}^2)$ is surface density \citep{Mo2010}.
The diffusion coefficient and timescale are estimated
in a standard manner based on the Larmor radius $R_L$
of a proton and the
coherence length $l_0$
of interstellar magnetic field assuming Kolmogorov-type turbulence.
The diffusion coefficient is formulated as
\begin{equation}
D=
\begin{cases}
\frac{cl_0}{3}\left[\left(\frac{R_L}{l_0}\right)^\frac{1}{3}+\left(\frac{R_L}{l_0}\right)^2\right] \quad&(R_L\leq\sqrt{H_{\rm g}l_0}) \\
\frac{cH_g}{3} &(R_L>\sqrt{H_{\rm g}l_0})
\end{cases}
\end{equation}
\citep{Aloisio2004}.  We assume $l_0=\min(30\,{\rm pc},H_{\rm g})$.
Magnetic field strength is estimated by equipartition with the ISM
energy density provided by star formation activity: $B^2/(8\pi)=\eta
E_{\rm SN}r_{\rm SN}t_{\rm adv}/V$, where $E_{\rm SN}$ is the kinetic
energy of a supernova, $r_{\rm SN}$ the event rate of supernova, and
$V=2\pi R_{\rm eff}^2H_{\rm g}$.  The parameter $\eta$ is determined
so that it reproduces the observed galactic magnetic field strength
$B=6\,{\rm \mu G}$ for MW \citep{Beck2008}.
From those estimations, the
gamma-ray luminosity and spectrum of a model galaxy are obtained.

Absorption of gamma-rays by EBL is taken into account in our model calculations, by
using the optical depth model of \citet{Inoue2013}. We do not
consider absorption within a galaxy in our calculation, because
these do not have large effects on our results for normal SFGs up to
a few TeV.  Pair-production optical depth becomes about unity around
5 TeV or higher for NGC 253 and M82 \citep{Inoue2011,Lacki2013}, and
photons beyond this energy may be suppressed compared with our model
predictions.  Pair-production optical depth in Arp 220 (an
ultra-luminous infrared galaxy) is even larger than those for NGC
253 and M82 \citep{Lacki2013}, but no galaxy in our sample is as
luminous in infrared as Arp 220.  Absorbed gamma-rays produce
secondary cascade emission by reprocessed photons at a lower energy
band, but this effect is ignored in our calculation, because this
component is not large in the TeV band compared with direct emission
\citep{Inoue2011}.

The overall normalization parameter $C$ is determined by fitting to
the observed gamma-ray luminosities of nearby galaxies, which are
listed in Table~\ref{tab:fit-parameter}.  This is the only free
parameter in this fitting, and the resultant value is $ C =
2.1\times10^{45}\,{\rm s^{-1}\,erg^{-1}}$ when we assume ${
  \Gamma_{\rm inj}=2.2}$ (see below). This value is reasonable
  compared with a theoretical expectation, which is $C = 3.2
  \times10^{45}\,{\rm s^{-1}\,erg^{-1}}$ when we assume the Salpeter initial mass function
  (IMF), a supernova mass threshold of $8 M_\odot$, 10\% of supernova
  explosion energy ($10^{51}$ erg) going into cosmic-rays, and
  $\Gamma_{\rm inj} = 2.2$ above 10 GeV. The sample of Table
~\ref{tab:fit-parameter} includes all star-forming galaxies detected
by {\it Fermi}-LAT at the time of S18.  Note that galaxies whose AGN might
contribute to gamma-ray emission are removed
\citep{Ackermann2012,YoastHull2017,Xi2020}, and that M31 is also
removed because its gamma-ray emitting region is different from gas
distribution, implying that the interaction between cosmic rays and
ISM is not the main source of gamma rays from this galaxy
\citep{Ackermann2017}.  After S18, gamma-ray detection has been
reported for M33 \citep{Ajello2020,Xi2020b,Xi2020}, along with a few
more galaxies whose gamma rays might originate from AGN activities. We
do not include this galaxy in the fitting, but the gamma-ray
luminosity predicted by our model will be compared with these
observations later.

In this paper, some physical parameters of galaxies in
Table~\ref{tab:fit-parameter} are modified from those in S18. For the
consistency with the KINGFISH catalog (see \S \ref{section:sample}),
SFR, gas mass, and stellar mass except for those of MW are derived in
the same way as \citet{RemyRuyer2014,RemyRuyer2015}.  SFRs are derived
from H$\alpha$ and IR luminosities using eq. (11) in
\citet{Kennicutt2009}, and IR luminosities are estimated from eq.  (5)
in \citet{Dale2002} using the IRAS data \citep{Sanders2003}.  Gas mass
is the sum of neutral and molecular hydrogen masses. 
Neutral hydrogen
mass is derived from 21-cm line flux, and H$_2$ mass is derived from
CO(1-0) line flux assuming a metallicity-dependent conversion factor
($X_\mathrm{CO-H_2}\propto Z^{-2}$) \citep{RemyRuyer2014}. 
See e.g. \citet{Obreschkow2009} for equations of gas mass derivation.
Stellar mass is estimated using
the equation derived in \citet{Eskew2012} using the Spitzer data
\citep{Dale2009}.

\begin{table*}
\begin{minipage}{\textwidth}
\renewcommand\thefootnote{\alph{footnote}}
\begin{center}
\caption{Physical properties of GeV-detected galaxies used to determine parameter $C$.}
\label{tab:fit-parameter}
\begin{tabular}{lcccccc}
\hline
Objects
&$D$\footnotemark[1]
&$L_\gamma$(0.1--800 GeV)\footnotemark[2]
&$\psi$\footnotemark[3]
&$M_\mathrm{gas}$\footnotemark[4]
&$M_\mathrm{star}$\footnotemark[5]
&$R_\mathrm{eff}$\footnotemark[6]\\
&(Mpc)
&($\mathrm{10^{39}\, erg\, s^{-1}})$
&$(\mathrm{M_\odot\,yr^{-1}})$
&$(10^9\,{\rm M_\odot})$
&$(10^9\,{\rm M_\odot})$
&$\mathrm{kpc}$
\\\hline\hline
MW & & $0.82\pm0.27$&2.6&4.9&50&6.0 \\
LMC & 0.05 & $0.032\pm0.001$ & 0.3 & 0.59 & 1.8 & 2.2\\
SMC & 0.06 & $0.0125\pm0.0005$ & 0.043 & 0.46 & 0.3 & 0.7\\
NGC 253 & 3.5 & $13\pm1$ & 3.3 & 3.2 & 54.4 & 0.5\\
M82 & 3.3 & $14.7\pm0.7$ & 4.4 & 4.7 & 21.9 & 0.3\\
NGC 2146 & 17.2 & $81.4\pm14.2$ & 11.4 & 10.4 & 87.1 & 1.7\\
\hline
\end{tabular}
\footnotetext[1]{Distances taken from
\citet{Kennicutt2008} for LMC, SMC and NGC 2146, \citet{Rekola2005} for NGC 253, and \citet{Foley2014} for M82.
}
\footnotetext[2]{Gamma-ray luminosities taken from
  \citet{Ajello2020} except for MW from 
\citet{Ackermann2012}.
Note that the energy range is 0.1-100 GeV for MW but 
0.1-800 GeV for all other galaxies. 
}
\footnotetext[3]{Star formation rates calculated from
\citet{Diehl2006,Makiya2011} for MW, and \citet{Sanders2003,Kennicutt2008,Kennicutt2009} for all other galaxies. See also text.
}
\footnotetext[4]{Total gas masses (atomic plus molecular hydrogens)
calculated from 
\citet{Paladini2007,RemyRuyer2014} for MW, \citet{StaveleySmith2003,Israel1997,Marble2010} for LMC, \citet{Stanimirovic1999,Israel1997,Marble2010} for SMC, \citet{Springob2005,Knudsen2007,Pilyugin2014} for NGC 253, \citet{Chynoweth2008,Weis2005,Moustakas2010} for M82, and \citet{RemyRuyer2014,Young1989} for NGC 2146. 
}
\footnotetext[5]{Stellar masses taken from 
\citet{BlandHawthorn2016} for MW, and \citet{Dale2009,Eskew2012} for other galaxies.
}
\footnotetext[6]{Effective radii taken from
\citet{Sofue2009} for MW, \citet{vanderMarel2006} for LMC, \citet{Gonidakis2009} for SMC, \citet{Strickland2004} for NGC 253 and M82, and \citet{Young1988} for NGC 2146.
}
\end{center}
\end{minipage}
\end{table*}

\begin{figure*}
\begin{minipage}{0.45\hsize}
\begin{center}
\includegraphics[width=\textwidth]{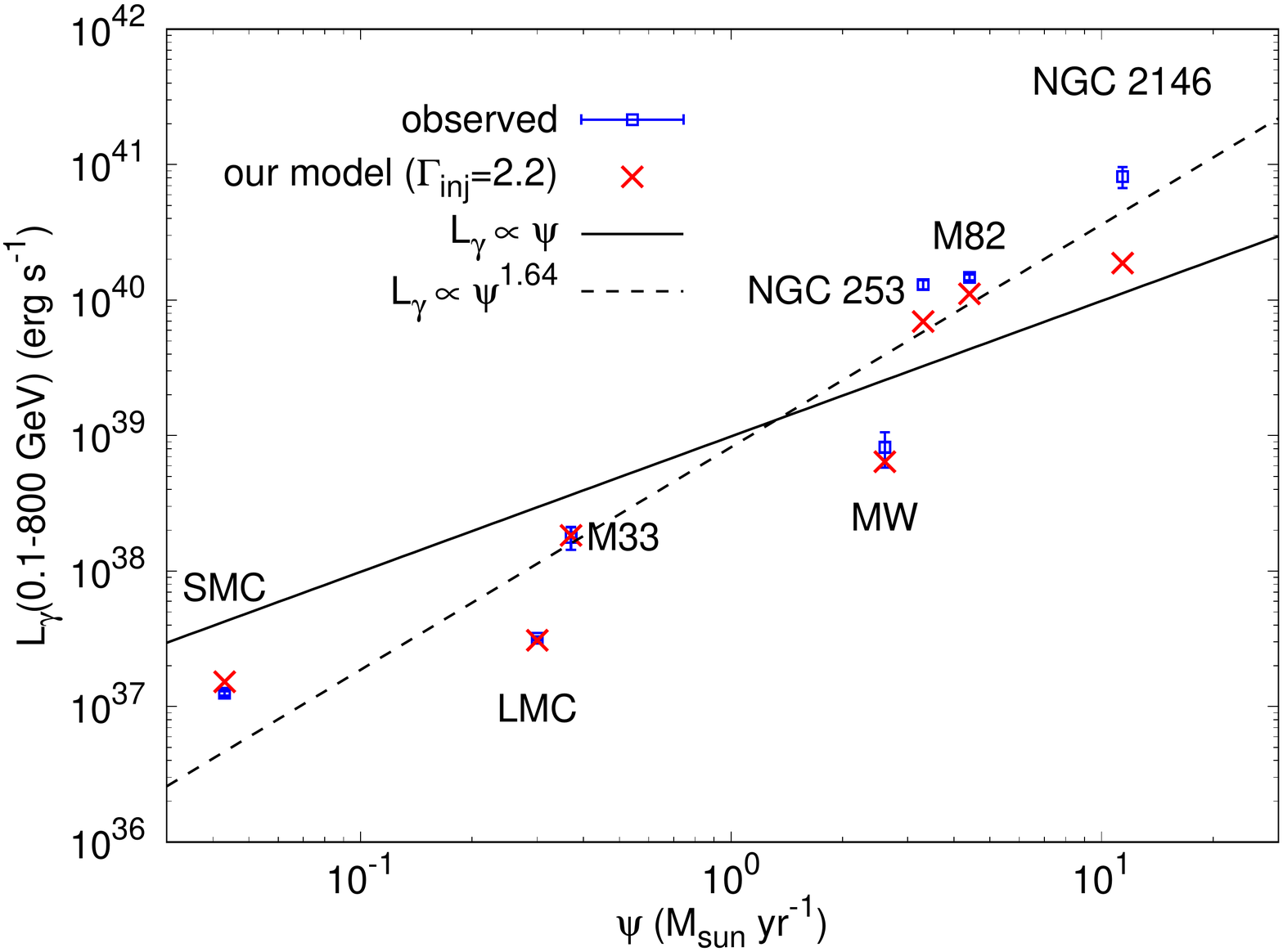}
\end{center}
\end{minipage}
\begin{minipage}{0.45\hsize}
\begin{center}
\includegraphics[width=\textwidth]{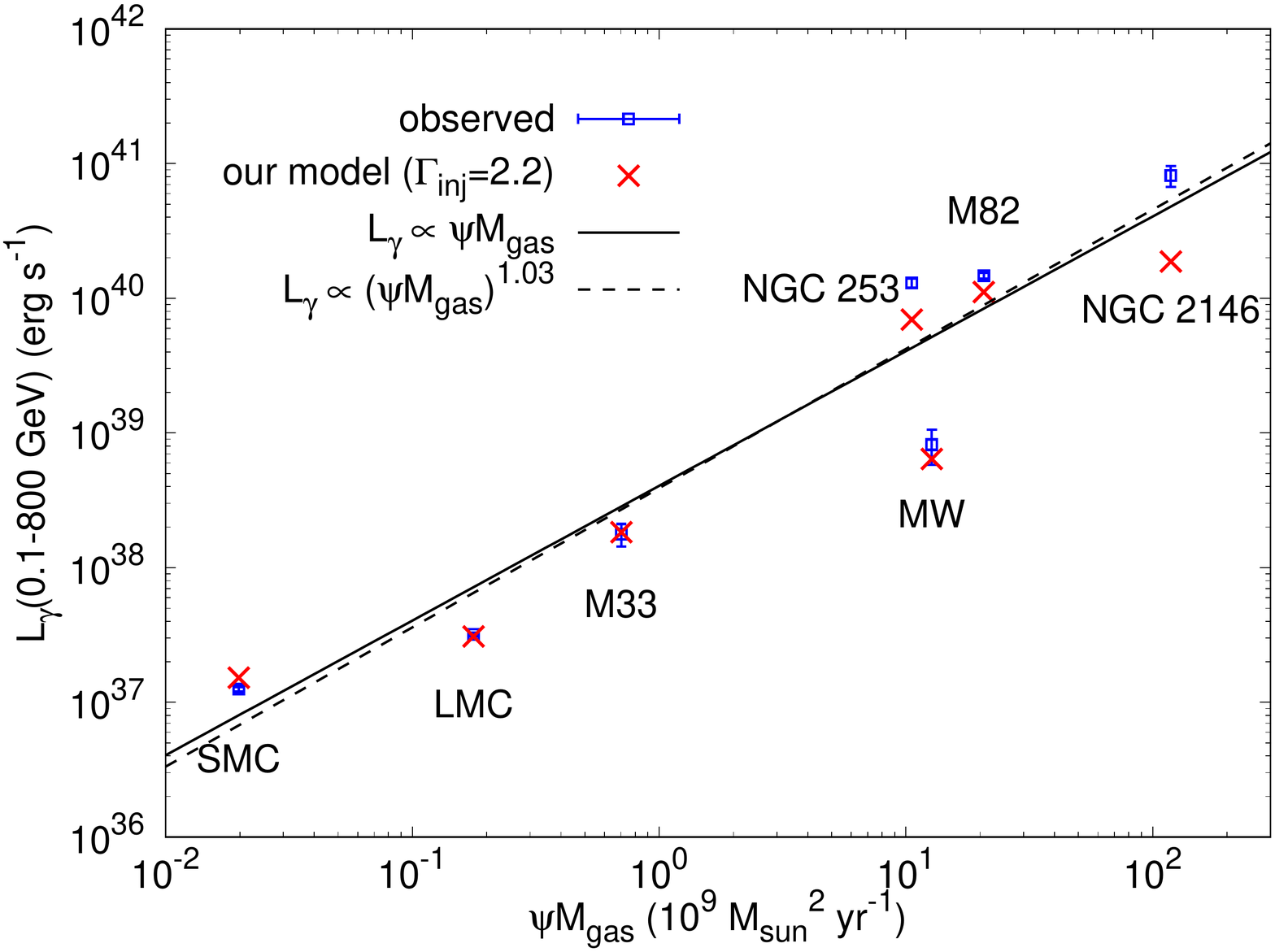}
\end{center}
\end{minipage}\\
\begin{minipage}{0.45\hsize}
\begin{center}
    \includegraphics[width=\textwidth]{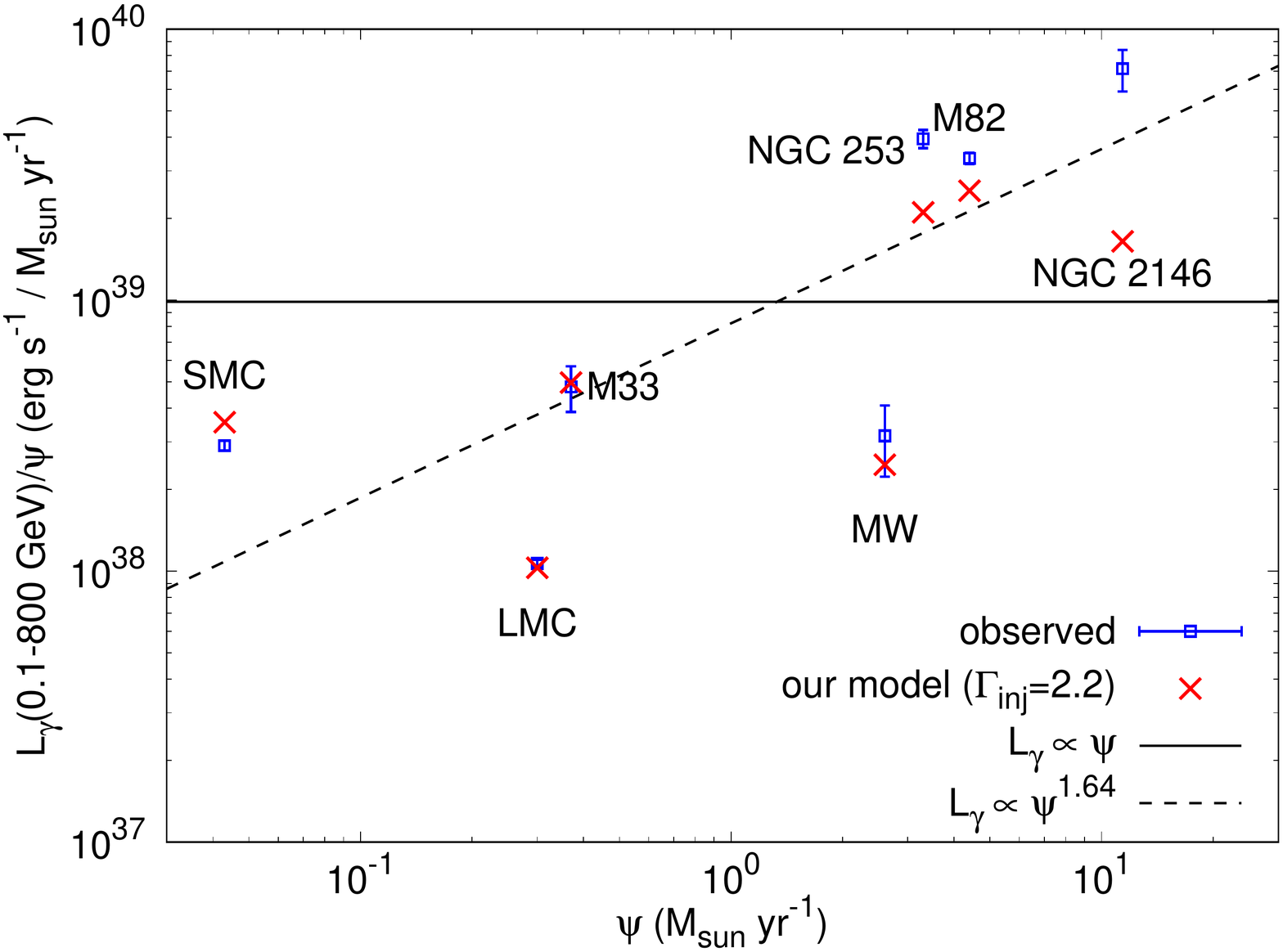}
\end{center}
\end{minipage}
\begin{minipage}{0.45\hsize}
\begin{center}
    \includegraphics[width=\textwidth]{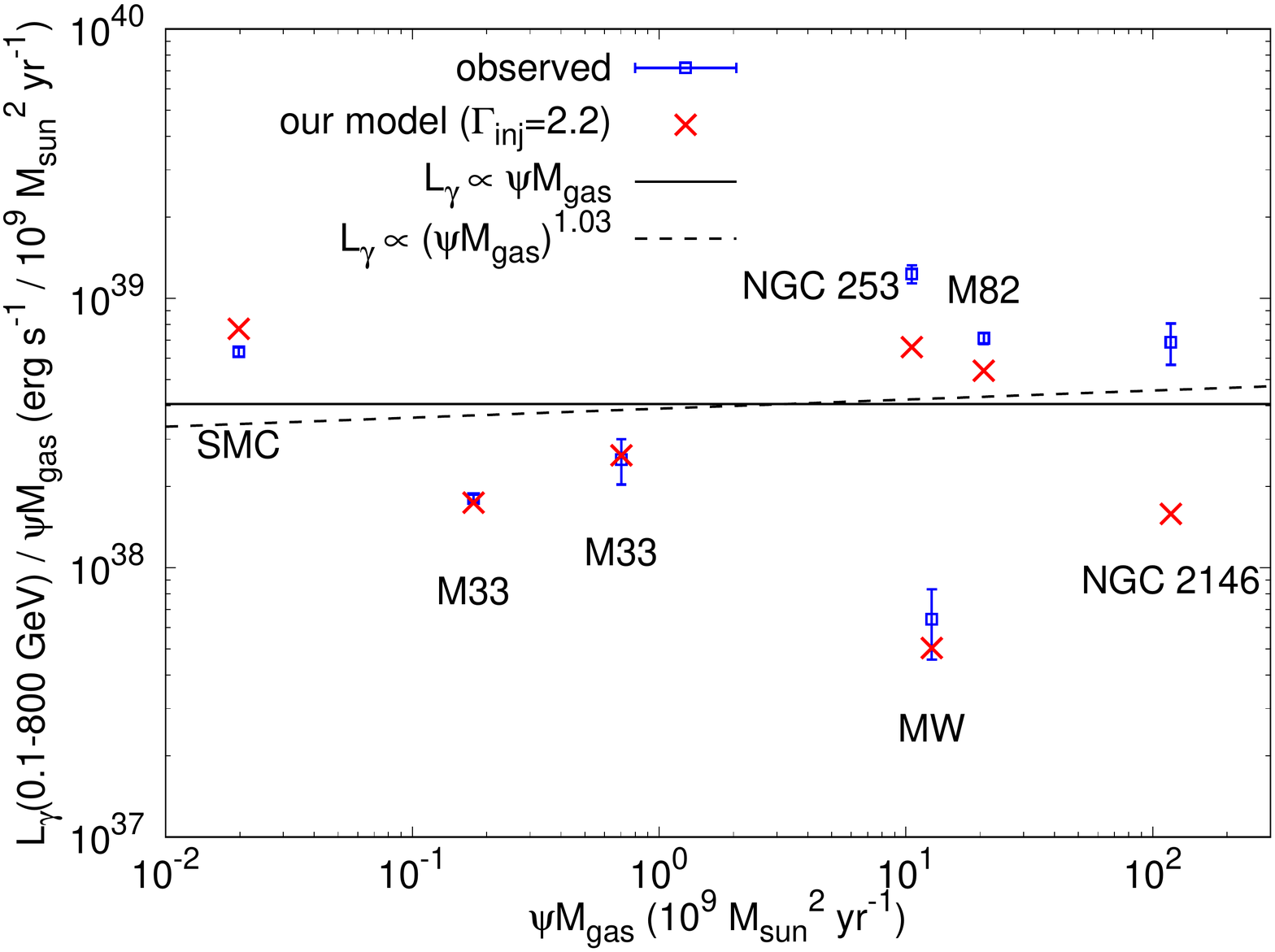}
\end{center}
\end{minipage}
\caption{Upper panels: gamma-ray luminosities of nearby 
  galaxies predicted by
  our model, in comparison with the observed luminosities. 
  The left panel is shown as a function of SFR, while
  the right panel as a function of SFR multiplied by gas mass.
  The solid lines are fits by a proportional relation,
  $L_\gamma$(0.1--800 GeV) $\propto \psi$ or $\psi M_{\rm gas}$, but the dashed
  lines are fits with a free power-law index. 
  It should be noted that the gamma-ray energy range is
  0.1--100 GeV for MW, while 0.1--800 GeV for all others. 
  Lower panels: the same as the upper panels, but the vertical
    axes are showing $L_\gamma/\psi$ and $L_\gamma/(\psi M_{\rm gas})$
    to show the deviation from the simple scalings to
    $\psi$ or $(\psi M_{\rm gas})$ more clearly. }
\label{fig:fitting}
\end{figure*}

Figure~\ref{fig:fitting} shows the comparison of our model and the
observed gamma-ray luminosities by {\it Fermi}-LAT.
Although NGC 2146 shows a slightly bad fit, 
our model
is in good agreement with observations as a whole,
and performs better than the
simpler assumptions such as $L_\gamma$(0.1--800 GeV) $\propto \psi$
(the calorimetric limit) or $L_\gamma$(0.1--800 GeV) $\propto \psi
M_{\rm gas}$ (intended to be the escape limit, though $M_{\rm gas}$
should be ISM density in reality).  
This is because we estimate the fraction of confined protons 
by considering more detailed physical conditions than simple models.
The data point for M33, which was
not used to determine $C$, is also shown here, and the luminosity
predicted by our model is in reasonable agreement with the newly
reported observed luminosity.
The observed luminosity of NGC 2146 is explained if we assume the calorimetric limit, which indicates that the emission region of this galaxy might be much smaller than our assumption, or perhaps there is another emission component.

\begin{figure*}
\begin{minipage}{0.45\hsize}
\begin{center}
\includegraphics[width=\linewidth]{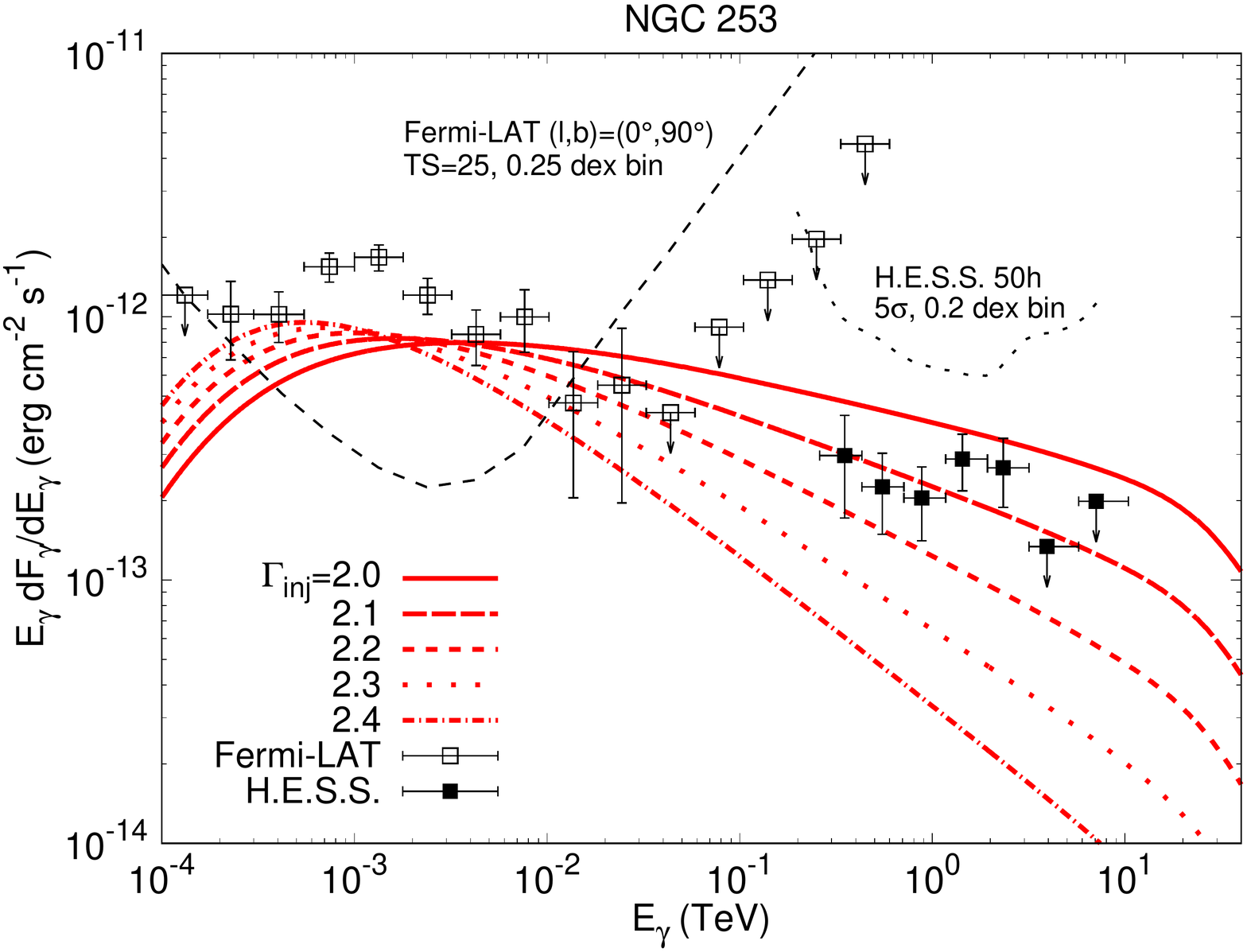}
\end{center}
\end{minipage}
\begin{minipage}{0.45\hsize}
\begin{center}
\includegraphics[width=\linewidth]{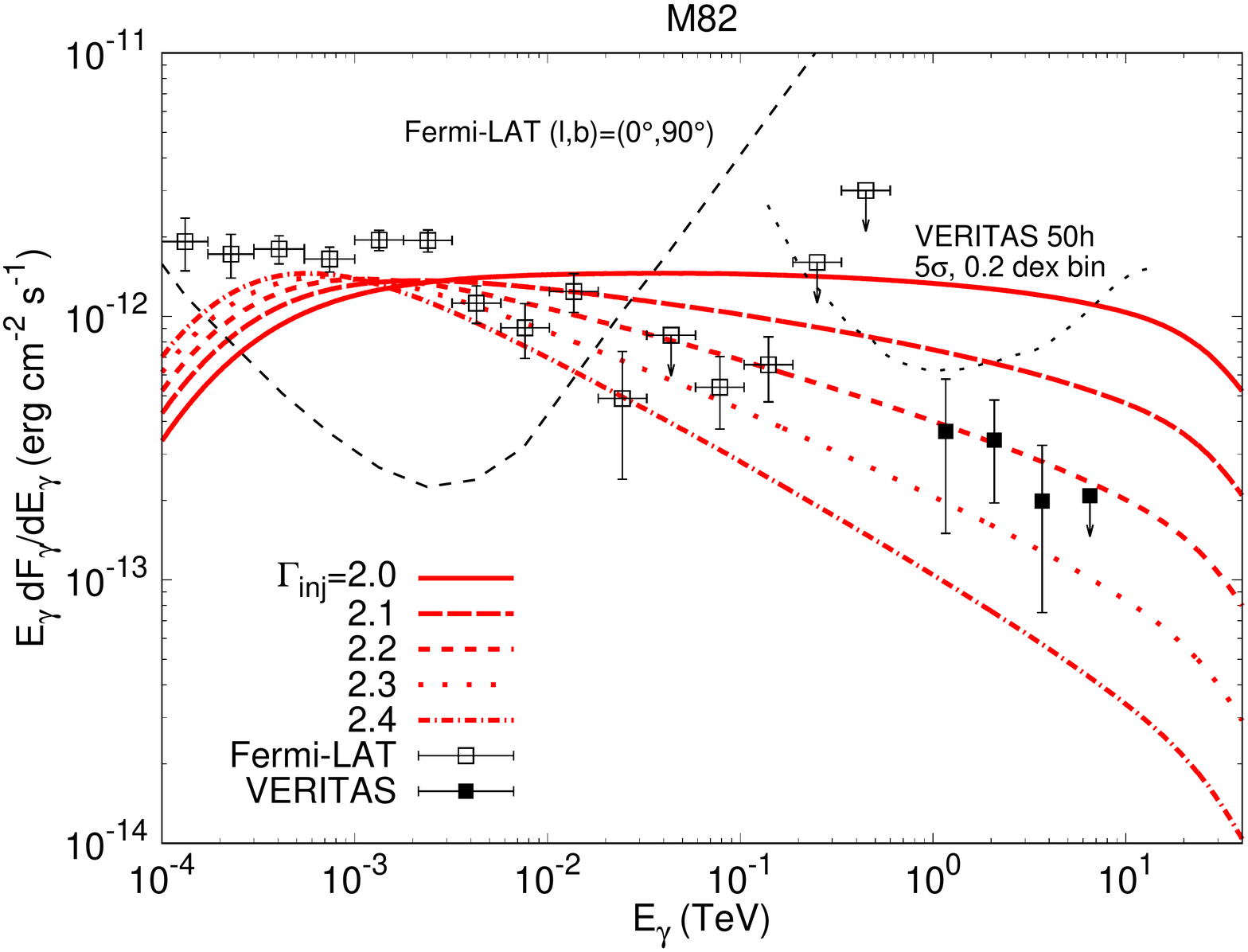}
\end{center}
\end{minipage}
\caption{GeV--TeV spectra of NGC 253 (left) and M82 (right) predicted
  by our model, in comparison with the observed spectra
  by \citealt{Ajello2020,H.E.S.S.Collaboration2018} for NGC 253 (10 yr
  for {\it Fermi}-LAT and 158 h for H.E.S.S) and
  \citealt{Ajello2020,VERITASCollaboration2009} for M82 (10 yr for
  {\it Fermi}-LAT and 137 h for VERITAS).  Sensitivities of {\it Fermi}-LAT
  (Pass 8 Release 3 Version 2, ${\rm TS}=25$ in a 0.25 dex photon
  energy bin) and TeV telescopes (5$\sigma$ detection by a 50 hr
  observation in a 0.2 dex photon energy bin) are also shown.
}
\label{fig:NGC253_and_M82}
\end{figure*}

Because NGC 253 and M82 have been detected in the TeV band, we show
our model spectra $E_\gamma \, dF_\gamma/dE_\gamma$, where
$dF_\gamma/dE_\gamma$ is differential energy flux per unit photon
energy, in comparison with the observed data of these galaxies in
Figure~\ref{fig:NGC253_and_M82} to verify that our model works not
only for GeV luminosity but also for GeV--TeV spectrum. 
In comparison, sensitivity curves of TeV telescopes
  \footnote{\citet{Holler2015}}
  \footnote{\url{https://veritas.sao.arizona.edu/about-veritas/veritas-specifications}} and {\it
    Fermi}-LAT\footnote{\url{http://www.slac.stanford.edu/exp/glast/groups/canda/lat_Performance.htm}}
    are shown.  The {\it Fermi}-LAT sensitivity depends on the source
    position on sky, and in this paper we adopt one of the four
    sensitivity curves at $(l, b) = (0^\circ, 0^\circ), (0^\circ,
    30^\circ), (0^\circ, 90^\circ)$ and $(45^\circ, 120^\circ)$. For
    each galaxy, we examine the sensitivity at its location using the
    sensitivity map of gamma-ray flux above 100 MeV, and choose the
    closest differential sensitivity curves from the four.  Our model
  is in good agreement with the observed spectra when we choose
  $\Gamma_\mathrm{inj} =$ 2.1--2.2.  It should be noted that our model
  spectrum for MW agrees with the observed spectrum of the diffuse
  Galactic gamma-ray background \citep{Ackermann2012b} when we set
  $\Gamma_\mathrm{inj} = 2.3$ (see Fig. 2 of S18).  This means that our model can be
  applied to a wide range of SFGs with a single spectral index
  $\Gamma_\mathrm{inj} \sim 2.2$ for the proton injection into the ISM.
  Note that the change of $\Gamma_{\rm inj}$ hardly affects the trend
seen in Fig. \ref{fig:fitting}, because it has a quantitatively
similar effect on all galaxies, provided that a single value is
used for all model galaxies.

\section{Sample}
\label{section:sample}

Physical parameters of the galaxy sample for this work are listed in
Table~\ref{tab:calc-parameter}.  This sample is constructed mainly
from the galaxies of KINGFISH project, which aims to understand the
physics of ISM in present-day galaxies \citep{Kennicutt2011}. 61
galaxies in the nearby universe (less than 30 Mpc) were chosen to
cover a wide range of galaxy properties and ISM environments, and
their SFR, gas mass, and stellar mass are compiled by
\citet{RemyRuyer2014,RemyRuyer2015}, making this sample suitable for
the purpose of this study.  However, we removed NGC 1377 and 1404 from
the sample, because SFR and/or gas mass are not available.  The total
number of KINGFISH galaxies used in this study is 59 (including NGC
2146, which is detected in the GeV band and listed in
Table~\ref{tab:fit-parameter}).

KINGFISH is not a complete sample about an observed flux, and some
nearby galaxies that are promising for detection by CTA may be
missed. Therefore we add galaxies that may be detected by CTA from the
IRAS Revised Bright Galaxy Sample (RGBS) \citep{Sanders2003} covering
all the sky. H$\alpha$ luminosities are available
\citep{Kennicutt2008} for RGBS galaxies that are bright enough
to be detected by CTA, allowing us to estimate SFR in the same way as
the KINGFISH galaxies. We then make a rough estimate of expected
gamma-ray luminosity as follows.  We define the value $L_\gamma(x)$ as
\begin{equation}
L_\gamma(x)=\left.E_\gamma\frac{dL_\gamma}{dE_\gamma}\right|_{x}\ ,
\end{equation}
where $dL_\gamma/dE_\gamma$ is differential luminosity per unit photon
energy.
If we assume the same $L_\gamma(1\, {\rm TeV})/{\rm SFR}$
ratio as NGC 253, we can relate TeV gamma-ray luminosities and SFRs as
\begin{equation}\label{eq:TeV}
  \frac{L_\gamma(1\,\mathrm{TeV})}{\mathrm{erg/s}}\sim9.1\times10^{37}
  \frac{\psi}{\mathrm{M_\odot yr^{-1}}} \ .
\end{equation}
The coefficient of this equation does not change significantly (within
20 \% change) if we use M82 instead of NGC 253. However, there may
be galaxies that have higher $L_\gamma(1\,{\rm TeV})/{\rm SFR}$ ratios
than NGC 253. In the most optimistic scenario where cosmic-ray protons
lose all the energy to {\it pp} collisions, this ratio
becomes $2.5\times10^{38}$ assuming the Salpeter IMF, 
the supernova mass threshold of $8 M_\odot$, cosmic-ray
energy of $10^{50}$ erg from a supernova, and a gamma-ray spectral
index of 2.2 above 0.1 GeV.

Using this optimistic ${ L_\gamma(1\,{\rm TeV})/{\rm SFR}}$ ratio to
estimate TeV gamma-ray luminosities, we selected galaxies that are not
in the KINGFISH catalog but whose expected TeV gamma-ray flux is
brighter than the CTA sensitivity limit at 1 TeV
\footnote{\url{https://www.cta-observatory.org/science/cta-performance/}}.
We then removed galaxies that are clearly AGNs or showing evidence of
AGN activity (NGC 1365, NGC 1068, NGC 4945, Cen A, and Arp 299).  The
famous starburst galaxy Arp 220, which also shows evidence of AGN
activity, is slightly below the CTA sensitivity limit by this
calculation. It should be noted that NGC 1068, NGC 4945, Arp 220, and
Arp 299 have been detected by {\it Fermi}-LAT and discussed in
\citet{Ajello2020}, but these are not included in our sample because we
want to select galaxies whose gamma-ray luminosities are purely by
star-forming activity. 
In Appendix~\ref{section:appendix}, we show spectra of M31 (removed due to its gas distribution; see Section~\ref{section:model}), NGC 1068, and Arp 220 to 
see how our model works for the galaxies removed from our sample. 
We then add the eight remaining galaxies to our
sample to ensure that it definitely contains all galaxies that are
potentially detectable by CTA.  SFR, gas mass, and stellar mass of
these galaxies are derived in the same way as those of KINGFISH, and
references for the necessary observed data are given in Table
\ref{tab:calc-parameter}.

Among those added eight galaxies, M33 and NGC 2403 have been already
detected by {\it Fermi}-LAT \citep{Xi2020,Ajello2020}. Especially, NGC
2403 is interesting because its emission is decaying with time and
this emission is coincident with SN 2004dj. However, time variability
is rather marginal, and to examine whether this emission is really
coming from a supernova, the estimate of steady gamma-ray flux from
its star-forming activity is important.

We need effective radii of galaxies to calculate gamma-ray
luminosities and spectra. We use the J-band half-light radii of the
2MASS Large Galaxy Atlas \citep{Jarrett2003} for most galaxies in our
sample.  Effective radii of dwarf galaxies DDO 53, Ho I, and M81dwB
are not available in the 2MASS data, and we use the results of
\citet{Oh2011} for these galaxies.

\begin{table*}
\begin{minipage}{\textwidth}
\renewcommand\thefootnote{\fnsymbol{footnote}}
\begin{center}
\caption{Physical parameters of galaxies in the sample of this
  work. The full table is available in supplementary material (online).}
\label{tab:calc-parameter}
\begin{tabular}{ccccccccccc}\hline
name &
$D\medspace (\mathrm{Mpc})$&ref&
$\psi (\mathrm{M_\odot\,yr^{-1}})$&ref&
$M_\mathrm{gas}\medspace (10^9\medspace \mathrm{M_\odot})$&ref&
$M_\mathrm{star}\medspace (10^9\medspace \mathrm{M_\odot})$&ref&
$R_\mathrm{eff}\medspace (\mathrm{kpc})$&ref\\
\hline\hline
NGC 5236 & 4.5 & (1) & 3 & (2)(3) & 17 & (4) & 49 & (5) & 3 & (6)\\
M33 & 0.84 & (2) & 0.37 & (2)(3) & 1.9 & (7)(8)(9) & 3.2 & (5) & 1.4 & (6)\\
NGC 6946 & 6.8 & (10) & 4.9 & (11) & 21 & (4) & 59 & (11) & 4.5 & (6)\\
IC 342 & 3.3 & (10) & 1.5 & (11) & 14 & (4) & 81 & (11) & 3.9 & (6)\\
NGC 7331 & 14 & (10) & 4.5 & (11) & 42 & (4) & 140 & (11) & 3.2 & (6)\\
... & ... & ... & ... & ... & ... & ... & ... & ... & ... & ...\\
\hline
\end{tabular}
\footnotetext{
References: (1) \citet{Galametz2011}, (2) \citet{Kennicutt2008}, (3) \citet{Sanders2003}, (4) \citet{RemyRuyer2014}, (5) \citet{Dale2009}, 
(6) \citet{Jarrett2003}, (7) \citet{Gratier2010}, (8) \citet{Heyer2004}, (9) \citet{Pilyugin2014}, (10) \citet{Kennicutt2011},
(11) \citet{RemyRuyer2015}, (12) \citet{Paturel2003}, (13) \citet{Allison2014}, (14) \citet{Israel1997}, (15) \citet{Leroy2008},
(16) \citet{Moustakas2010}, (17) \citet{Madden2013}, (18) \citet{Oh2011}
}
\end{center}
\end{minipage}
\end{table*}

\section{Result}
\label{section:result}
\subsection{TeV gamma-ray detectability by CTA}

\begin{figure*}
\begin{minipage}{0.45\hsize}
\begin{center}
\includegraphics[width=\textwidth]{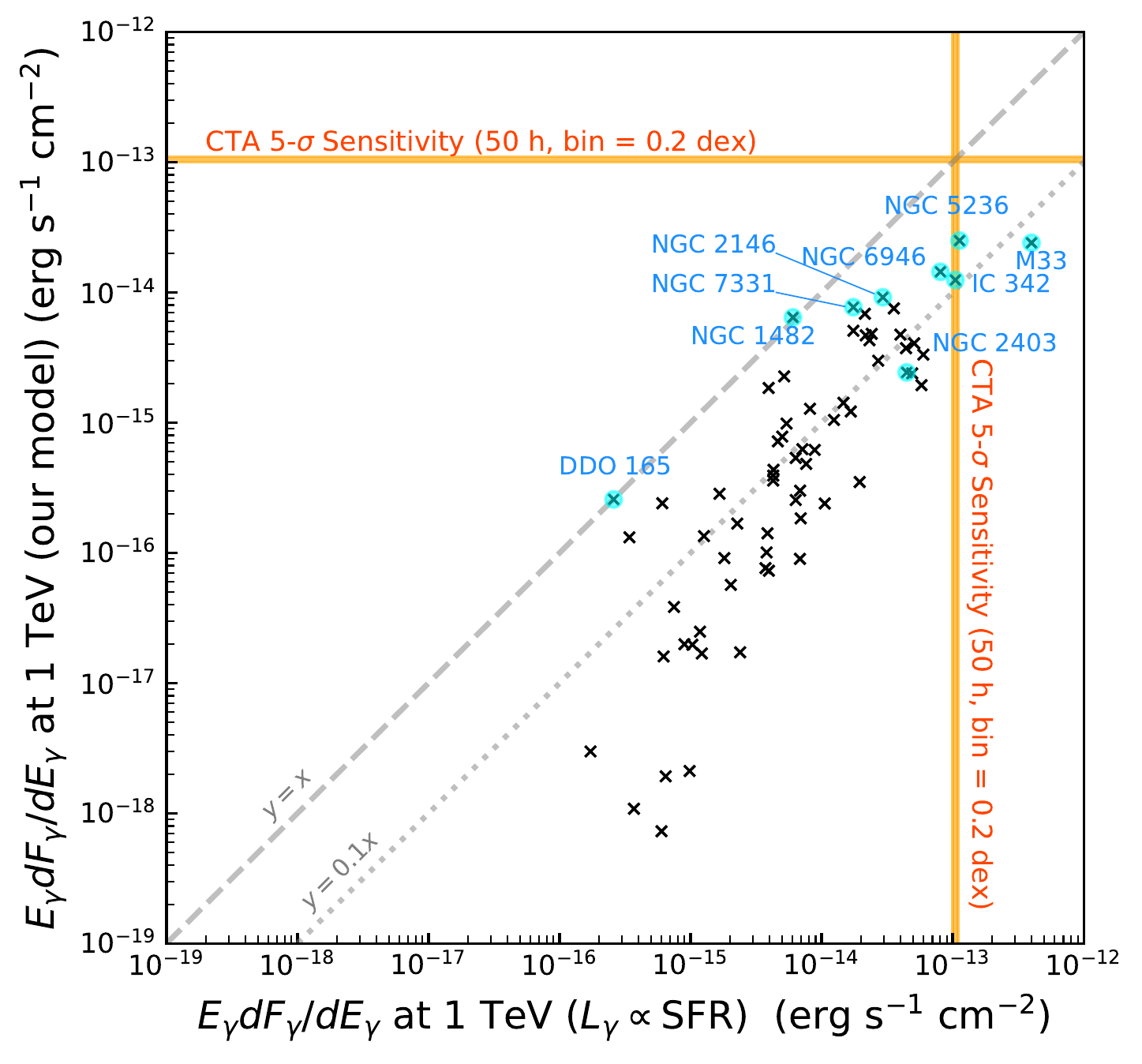}
\end{center}
\end{minipage}
\begin{minipage}{0.45\hsize}
\begin{center}
\includegraphics[width=\textwidth]{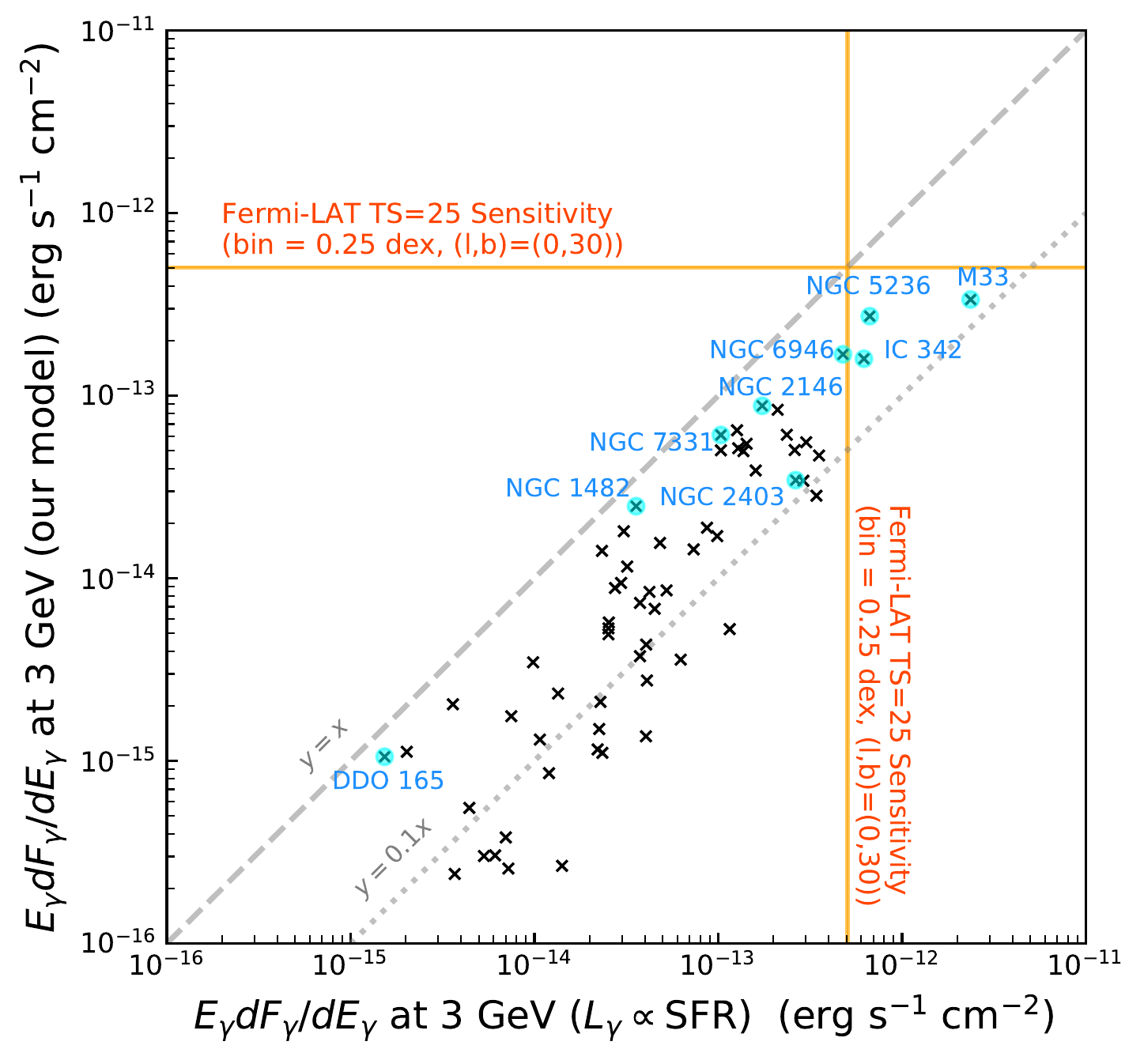}
\end{center}
\end{minipage}
\caption{The fluxes at $1\medspace \mathrm{TeV}$ (left) and
  $3\,\mathrm{GeV}$ (right) predicted by our model assuming
  $\Gamma_\mathrm{inj}=2.2$ are shown in the vertical axes, in
  comparison with the simpler prediction assuming $L_\gamma \propto$
  SFR and the $L_\gamma$/SFR ratio of NGC 253 shown in the horizontal
  axes.  
  Here we adopt the {\it
    Fermi}-LAT sensitivity at $(l, b) = (0^\circ, 30^\circ)$ that is
  intermediate among those of four sky locations provided at the {\it
    Fermi}-LAT website, and
    CTA South sensitivity. 
    Blue-highlighted labeled galaxies are discussed in the text or
      those with spectra shown in figures.
    }
\label{fig:flux-flux}
\end{figure*}

In Figure~\ref{fig:flux-flux} (left), we show the predicted fluxes at
1 TeV by our model assuming $\Gamma_\mathrm{inj}=2.2$. In the plot,
these fluxes are compared with the simple expectation using
Equation~\ref{eq:TeV}.  The fluxes predicted by our model are
typically 15 times lower than the simple estimates, indicating that
cosmic rays responsible for TeV emission are less confined than those
in NGC 253.

For comparison,
the same plot is made for within the {\it Fermi}-LAT energy band at 3 GeV
as the right panel of the same figure,
in comparison with the simple estimates being proportional
to SFR:
\begin{equation}
  \frac{L_\gamma(3\,\mathrm{GeV})}{\mathrm{erg/s}}\sim4.7\times10^{38}
  \frac{\psi}{\mathrm{M_\odot yr^{-1}}} \ ,
\end{equation}
where the coefficient is again determined based on NGC 253.
The fluxes predicted
by our model are again lower than the simple estimates by a factor of
7, but the differences are smaller than the TeV band (by a factor of
$15/7$), which is reasonable because of stronger confinement expected
for lower energy cosmic rays.
Note that M33 has been detected by {\it Fermi}-LAT as mentioned above,
  and \citet{Ackermann2012} reported a signature of detection (${\rm TS}\sim 15$) for NGC 5236. Other galaxies have not yet been detected by {\it Fermi}-LAT,
  which is consistent with our model prediction.

\begin{figure*}
\begin{minipage}{0.4\hsize}
\begin{center}
\includegraphics[width=\linewidth]{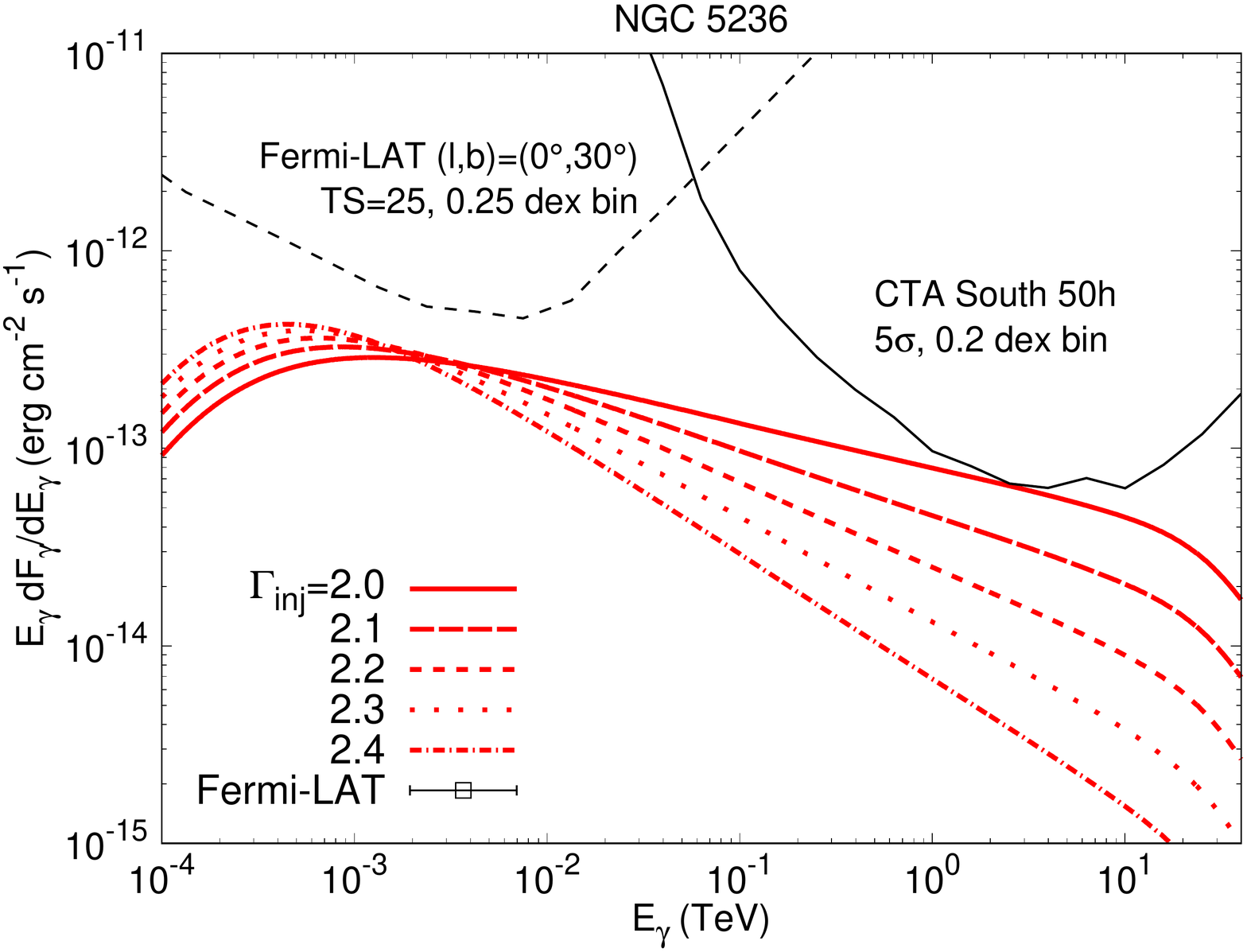}
\end{center}
\end{minipage}
\begin{minipage}{0.4\hsize}
\begin{center}
\includegraphics[width=\linewidth]{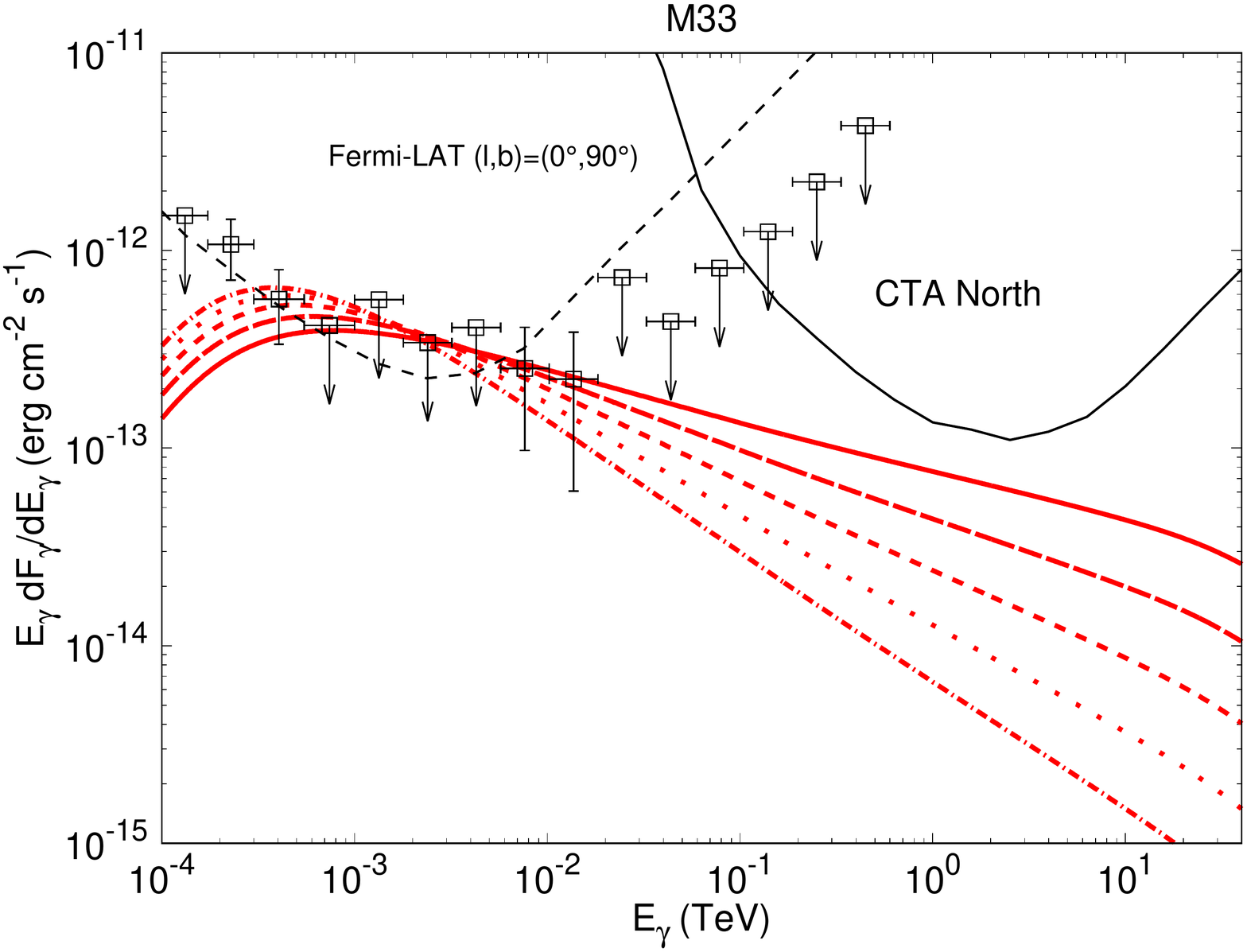}
\end{center}
\end{minipage}\\
\begin{minipage}{0.4\hsize}
\begin{center}
\includegraphics[width=\linewidth]{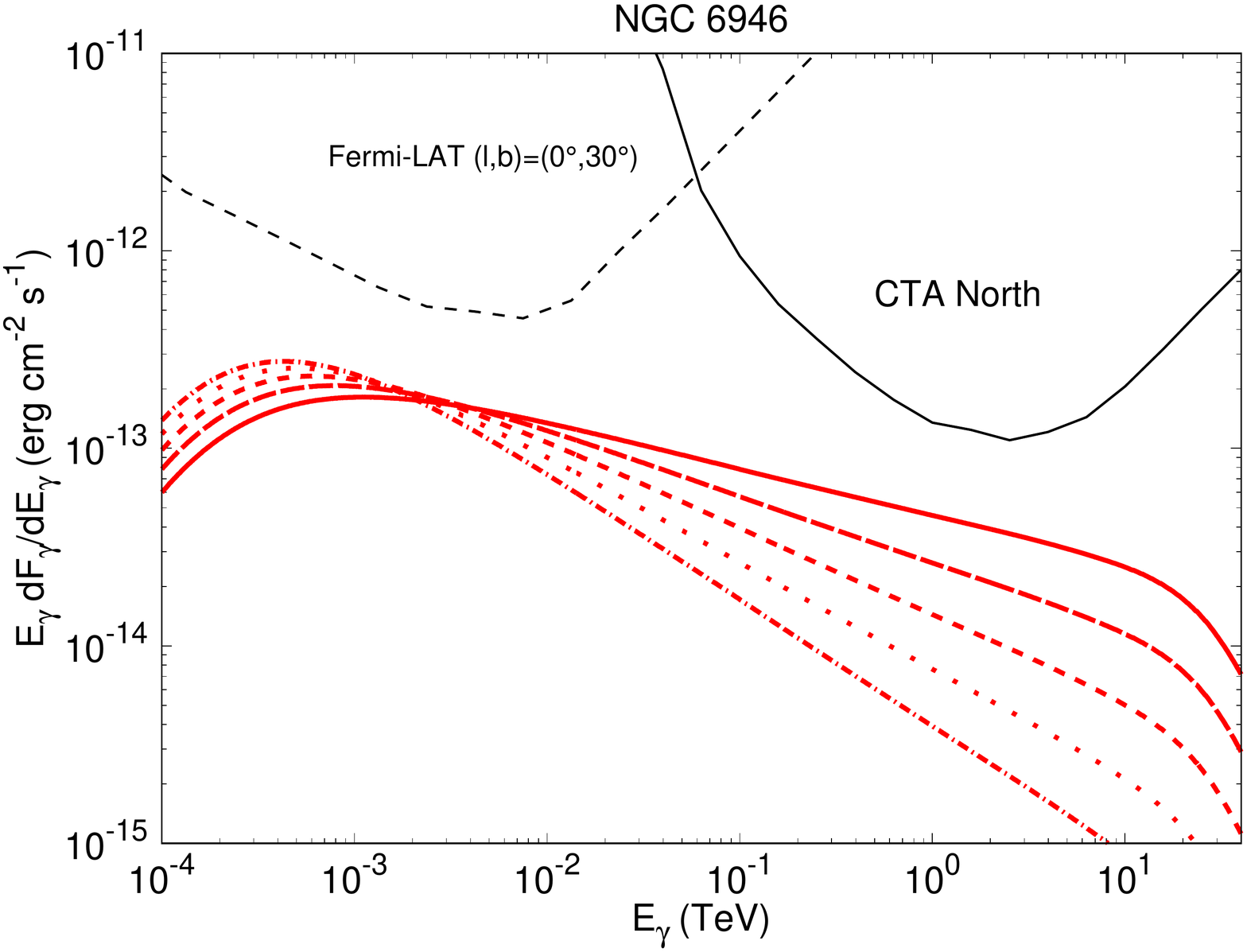}
\end{center}
\end{minipage}
\begin{minipage}{0.4\hsize}
\begin{center}
\includegraphics[width=\linewidth]{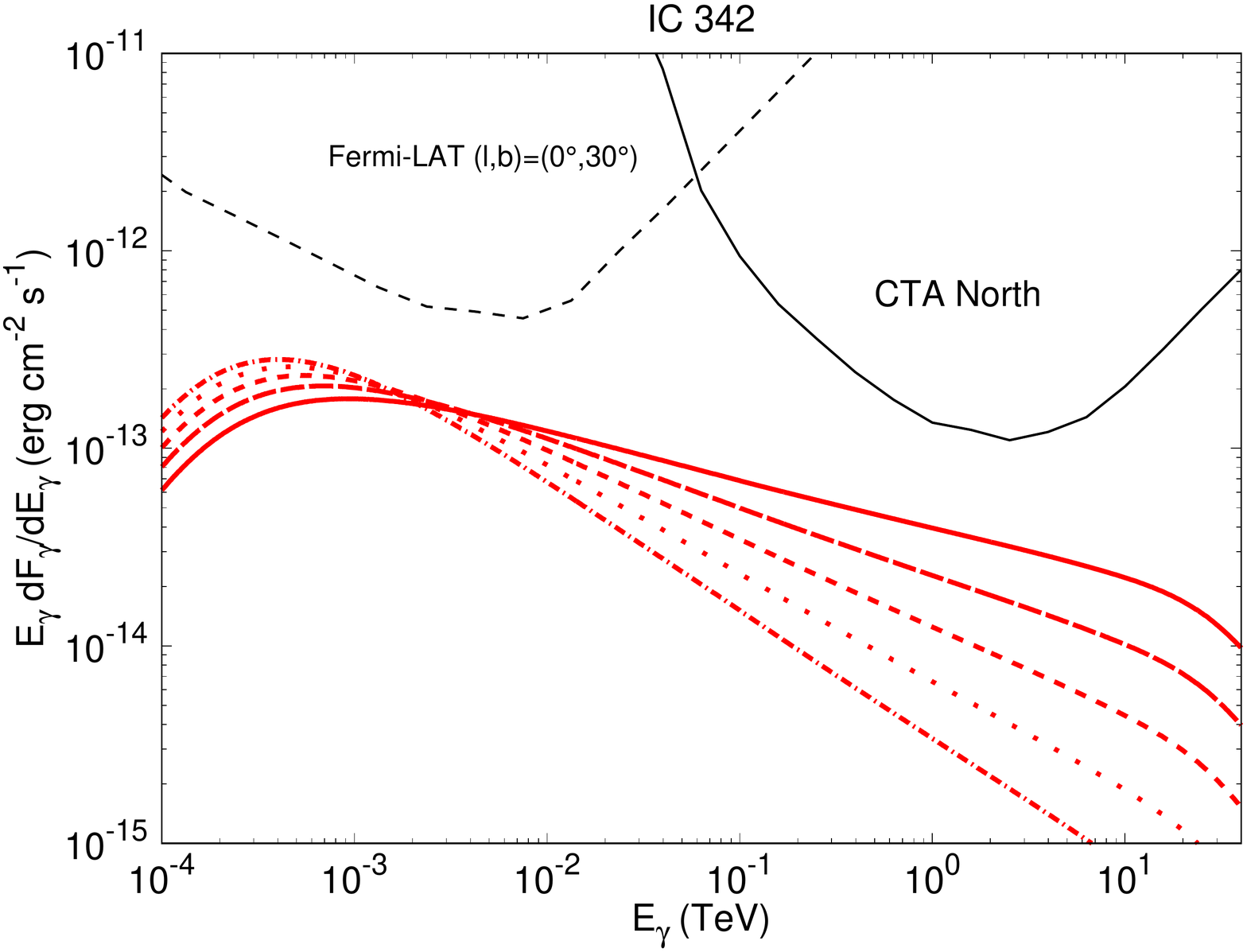}
\end{center}
\end{minipage}\\
\begin{minipage}{0.4\hsize}
\begin{center}
\includegraphics[width=\linewidth]{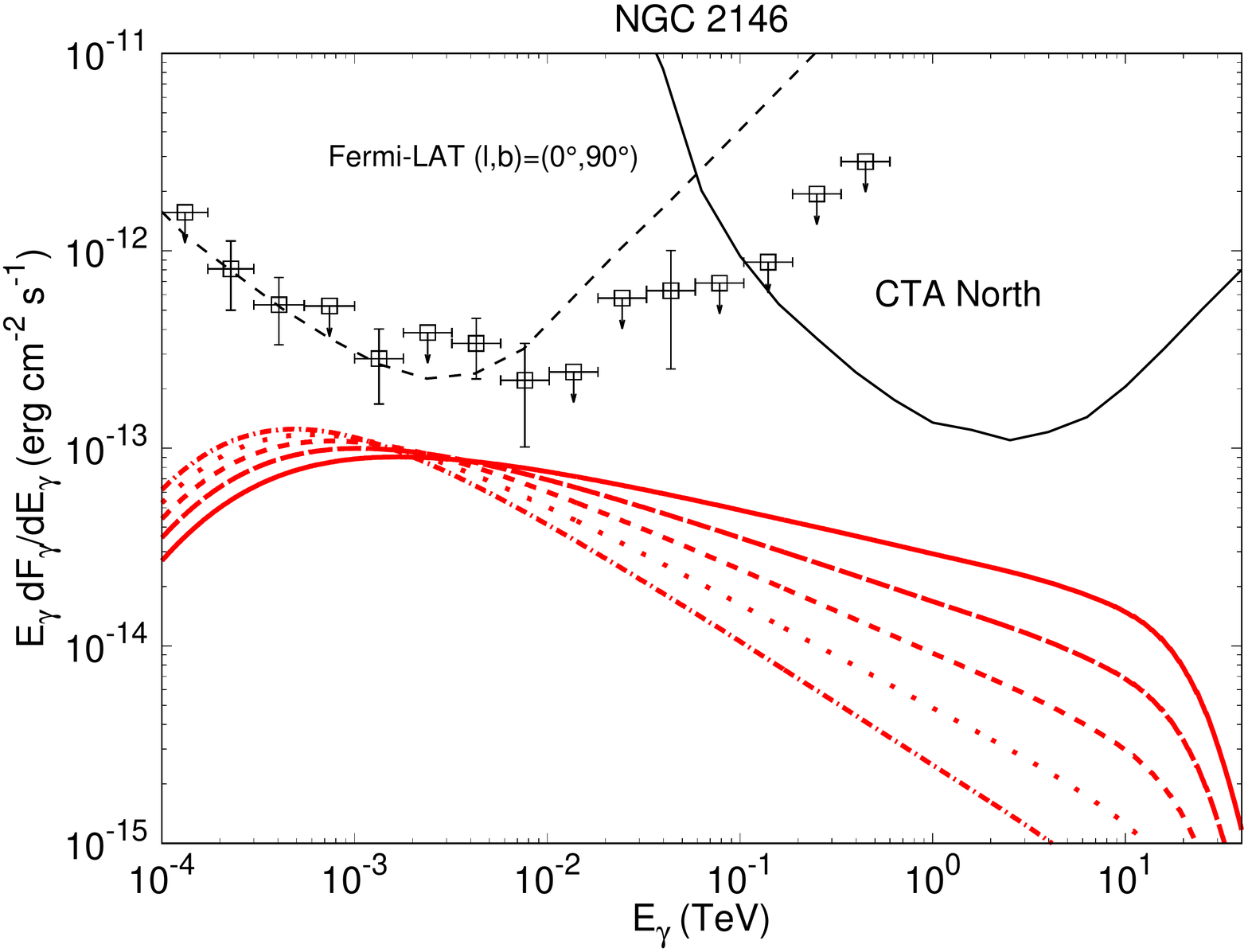}
\end{center}
\end{minipage}
\begin{minipage}{0.4\hsize}
\begin{center}
\includegraphics[width=\linewidth]{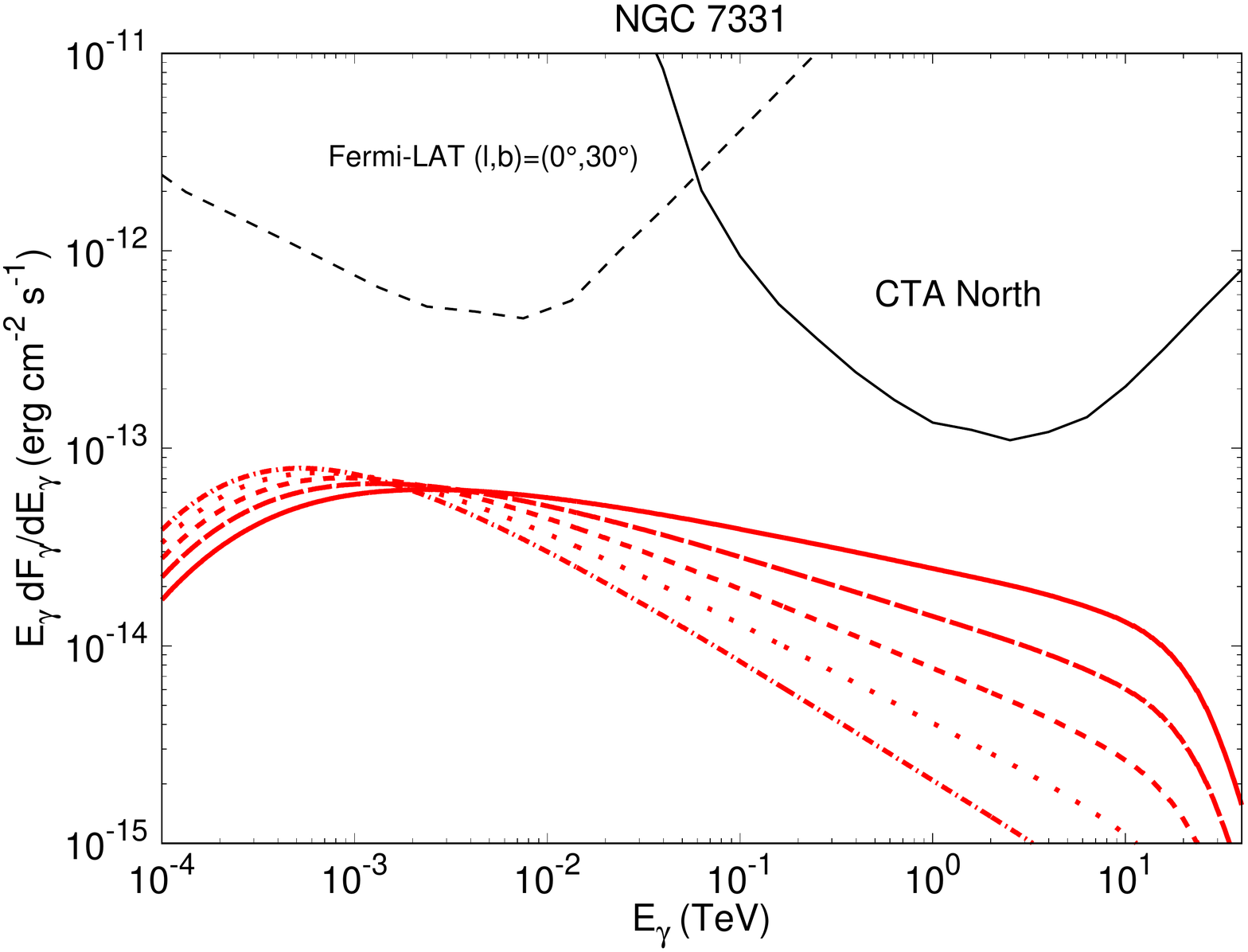}
\end{center}
\end{minipage}\\
\begin{minipage}{0.4\hsize}
\begin{center}
\includegraphics[width=\linewidth]{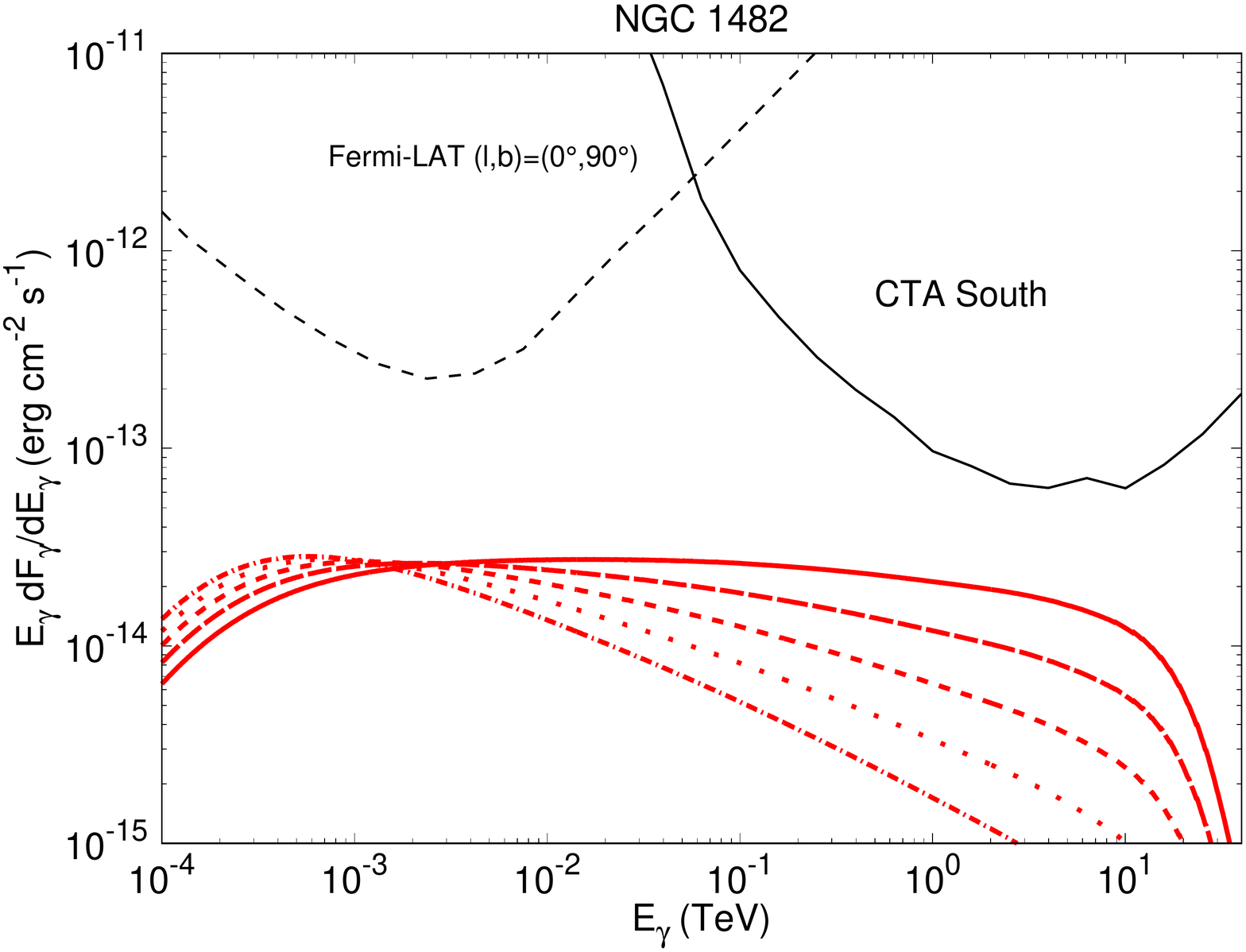}
\end{center}
\end{minipage}
\begin{minipage}{0.4\hsize}
\begin{center}
\includegraphics[width=\linewidth]{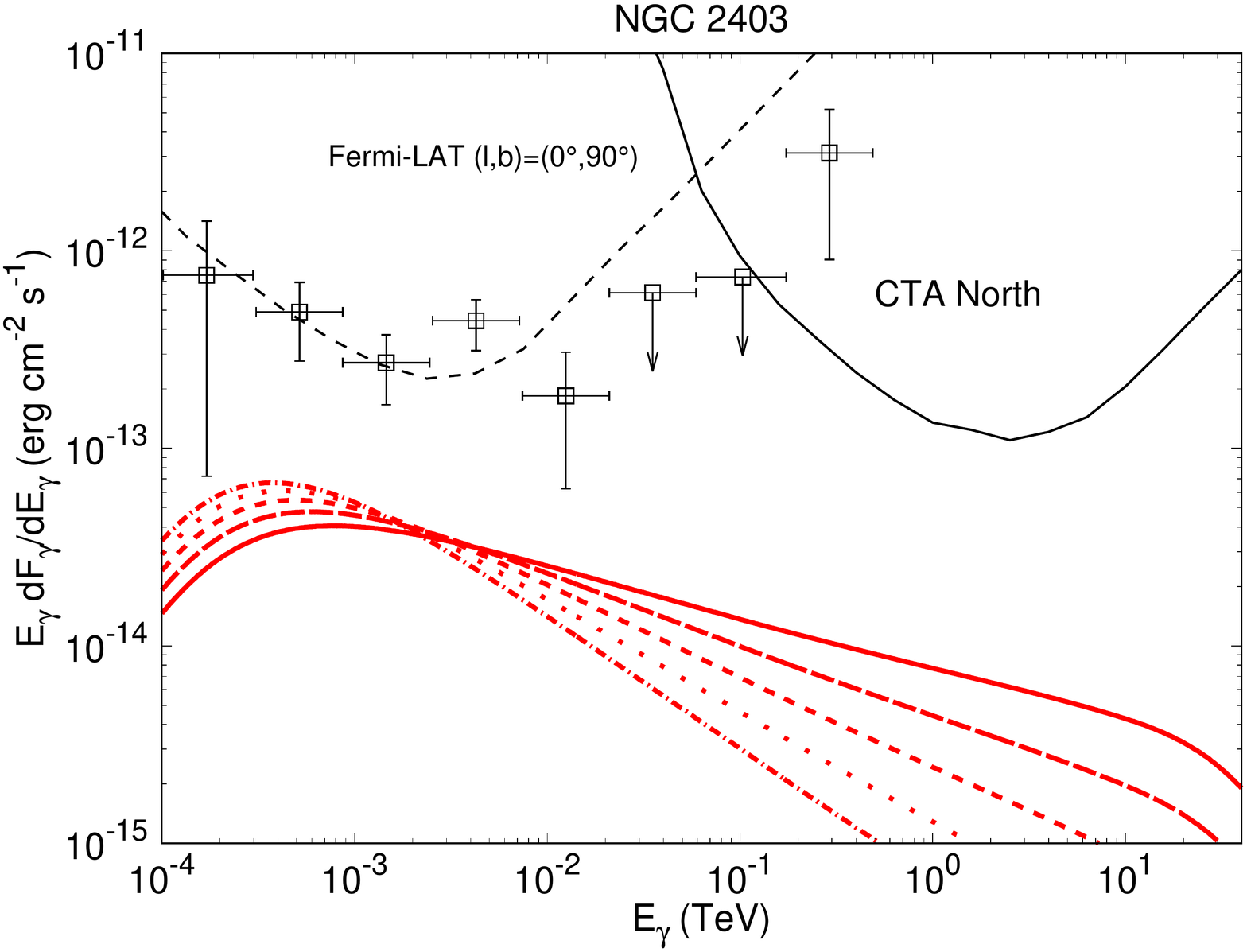}
\end{center}
\end{minipage}\\
\caption{
  Examples of gamma-ray spectra of nearby galaxies predicted by our
  model. 
  The sensitivities of CTA South or North are shown, depending on their declinations. 
  NGC 5236, M33, NGC 6946, IC 342, NGC 2146, and NGC 7331 are the
  brightest among the sample considered in this work (except those
  already detected in TeV).
  NGC 1482 is predicted to have
  a particularly hard spectrum which may be detected by CTA.  NGC 2403
  has been detected by {\it Fermi}-LAT, but the GeV flux 
  might originate from a supernova.  Data
  points are taken from \citet{Xi2020} for NGC 2403 (5.7 yr),and \citet{Ajello2020} (10 yr) for other galaxies. 
}
\label{fig:spectrum_examples}
\end{figure*}

Table~\ref{tab:calculated-flux} presents gamma-ray 
fluxes ($E_\gamma dF_\gamma/dE_\gamma$ at 1 TeV) 
predicted by our model for the galaxies considered in this paper,
in descending order when $\Gamma_\mathrm{inj}=2.2$.  
The spectra of
the top six galaxies (NGC 5236, M33, NGC 6946, IC 342, NGC 2146, and NGC 7331)
are shown in Figure~\ref{fig:spectrum_examples}. 
The fluxes of these
galaxies assuming $\Gamma_\mathrm{inj}=2.2$ are lower than the CTA
sensitivity limit (50h, 5$\sigma$) by a factor of a few.  It should be
noted that the sensitivity limit quoted here is in a photon energy bin
of a 0.2 dex width, and a fainter flux can be detected if a wider
energy bin width is adopted. In fact, NGC 253 has been detected by
H.E.S.S, whose flux is also fainter than its sensitivity limit by a
similar factor.  Therefore we conclude that there is a reasonable
chance of detection of these galaxies by CTA if sufficient
observing time is devoted. 
We note that the chance of TeV gamma-ray detection for NGC 2146 is enhanced if the TeV flux is higher than that of our model by the same factor as the discrepancy at GeV energy range.

\begin{table*}
\begin{center}
  \caption{List of gamma-ray fluxes ($E_\gamma \, dF_\gamma/dE_\gamma$ at
    1 TeV) predicted by our model, in units of
    $\mathrm{erg\,s^{-1}\,cm^{-2}}$ in the descending order when
    $\Gamma_{\rm inj} = 2.2$. The full table is available in
    supplementary material (online).  }
\label{tab:calculated-flux}
\begin{tabular}{|l|ccccc|}
\hline
name&$\Gamma_\mathrm{inj}=2.0$&$\Gamma_\mathrm{inj}=2.1$&$\Gamma_\mathrm{inj}=2.2$&$\Gamma_\mathrm{inj}=2.3$&$\Gamma_\mathrm{inj}=2.4$\\
\hline\hline

NGC 5236 & $7.9\times10^{-14}$ & $4.6\times10^{-14}$ & $2.5\times10^{-14}$ & $1.3\times10^{-14}$ & $6.8\times10^{-15}$\\
M33 & $7.6\times10^{-14}$ & $4.4\times10^{-14}$ & $2.4\times10^{-14}$ & $1.3\times10^{-14}$ & $6.6\times10^{-15}$\\
NGC 6946 & $4.6\times10^{-14}$ & $2.6\times10^{-14}$ & $1.4\times10^{-14}$ & $7.6\times10^{-15}$ & $3.9\times10^{-15}$\\
IC 342 & $3.9\times10^{-14}$ & $2.3\times10^{-14}$ & $1.2\times10^{-14}$ & $6.6\times10^{-15}$ & $3.4\times10^{-15}$\\
NGC 2146 & $2.9\times10^{-14}$ & $1.7\times10^{-14}$ & $9.2\times10^{-15}$ & $4.8\times10^{-15}$ & $2.5\times10^{-15}$\\
NGC 7331 & $2.5\times10^{-14}$ & $1.4\times10^{-14}$ & $7.7\times10^{-15}$ & $4.1\times10^{-15}$ & $2.1\times10^{-15}$\\
... & ... & ... & ... & ... & ...\\
\hline
\end{tabular}
\end{center}
\end{table*}

\subsection{Properties of TeV gamma-ray spectrum}

\begin{figure}
\begin{center}
\includegraphics[width=\columnwidth]{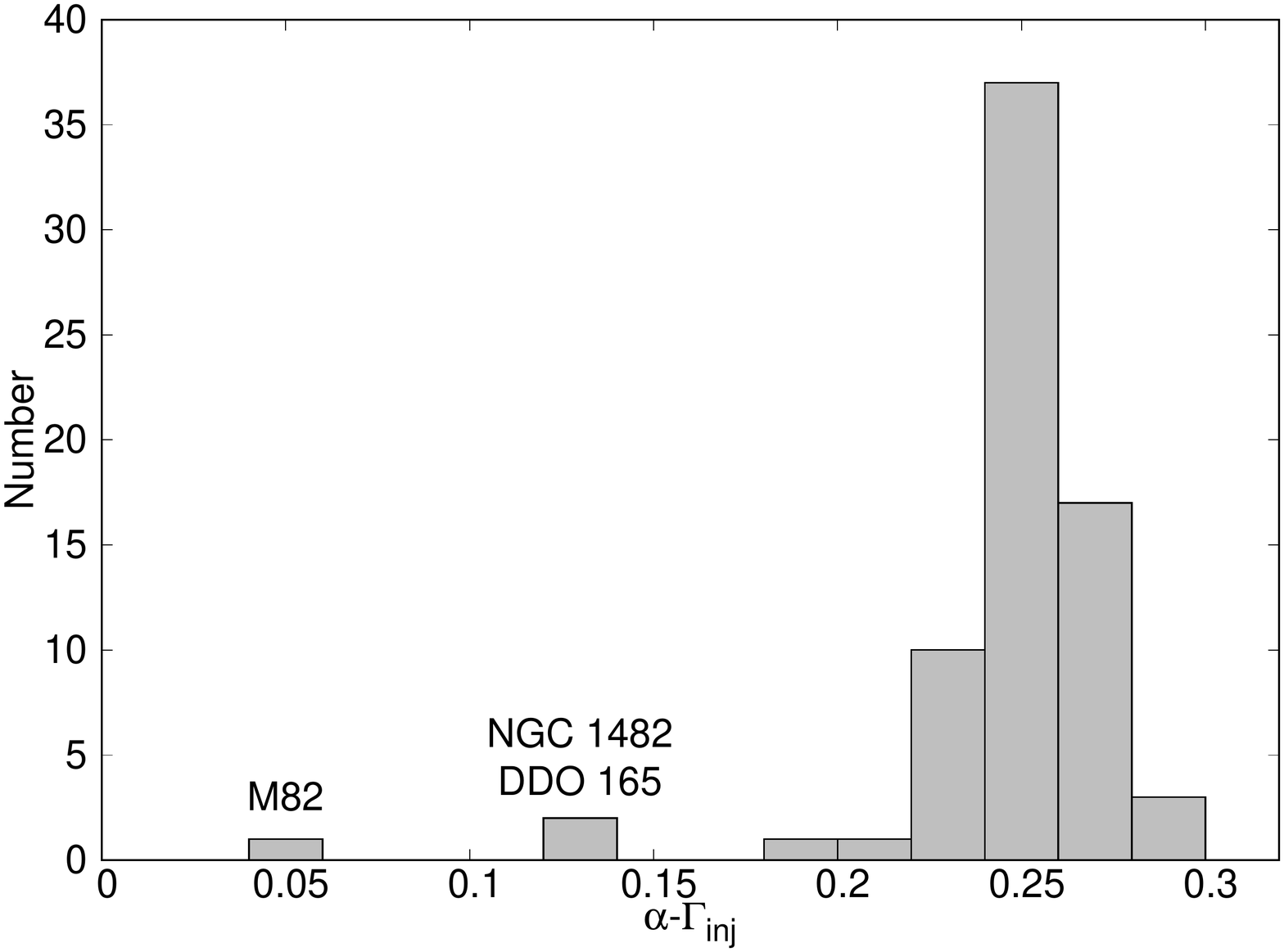}
\caption{Histogram of $\alpha-\Gamma_\mathrm{inj}$ assuming $\Gamma_\mathrm{inj}=2.2$, where $\alpha$ is the power-law index
  of a photon spectrum, obtained by a fit to { the modeled gamma-ray
  spectra of galaxies} in 0.1--10 TeV.
}
\label{fig:histgram}
\end{center}
\end{figure}

The power-law index $\Gamma_{\rm inj}$ of injected proton spectrum is
a free model parameter in our model, but the gamma-ray energy spectrum is
calculated by the model taking into account cosmic-ray propagation and
escape.  Then the difference between gamma-ray and proton spectra is a
good indicator of the efficiency of cosmic-ray collisions with ISM
before its escape. Therefore we show the histogram of $\alpha - \Gamma_{\rm
  inj}$ in Figure~\ref{fig:histgram}, where $\alpha$ is the photon
spectral index defined as $dF_\gamma/dE_\gamma \propto
E_\gamma^{-\alpha+1}$.  We calculated $\alpha$ by a power-law fit to
the modeled spectra in the range of 0.1--10 TeV.  In most galaxies
$\alpha - \Gamma_{\rm inj}$ is larger than 0.2, but M82, NGC 1482 and DDO
165 have notably smaller values than other galaxies, implying that
cosmic-ray interaction is particularly efficient in these.

\begin{figure*}
\begin{minipage}{0.45\hsize}
\begin{center}
\includegraphics[width=\linewidth]{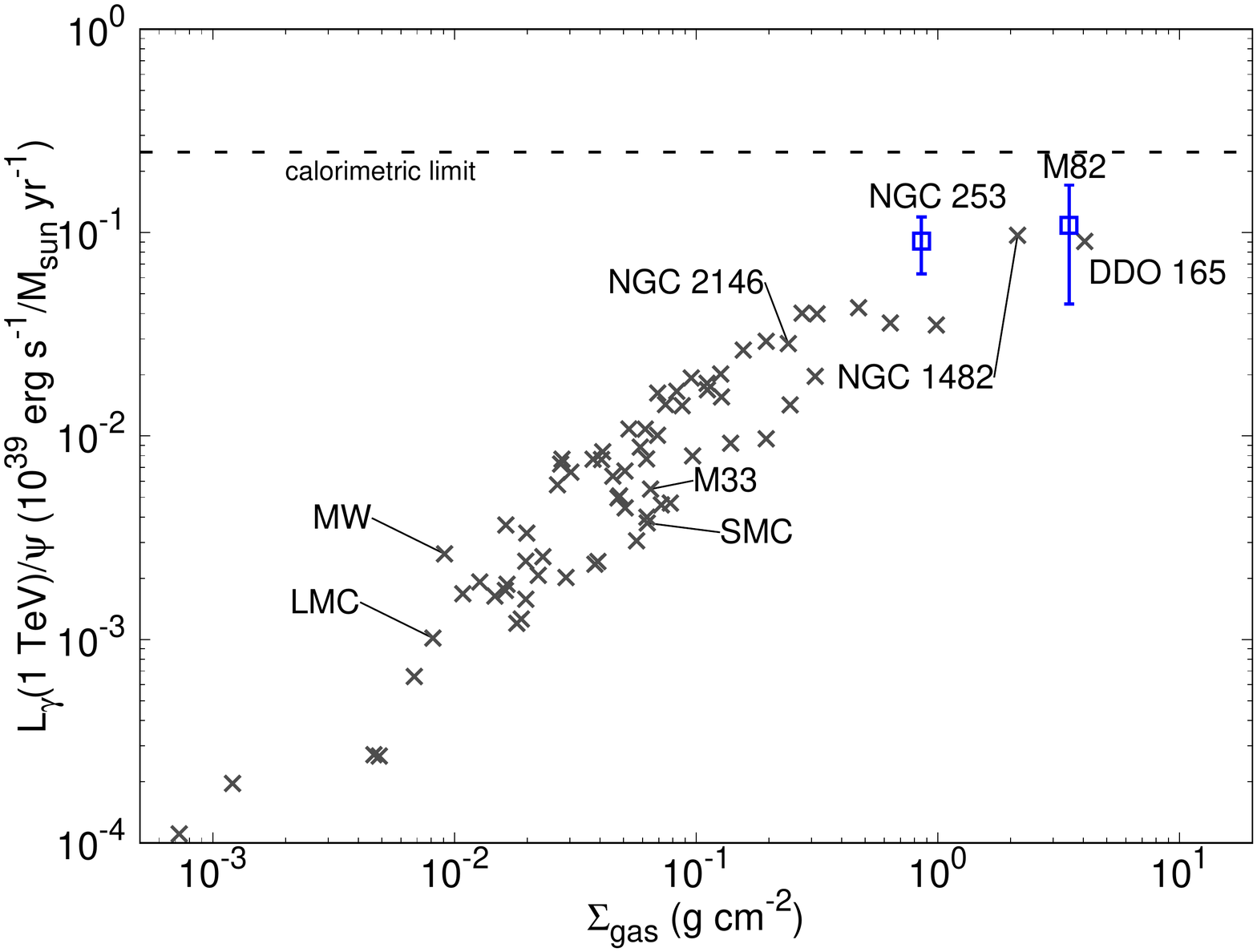}
\end{center}
\end{minipage}
\begin{minipage}{0.45\hsize}
\begin{center}
\includegraphics[width=\linewidth]{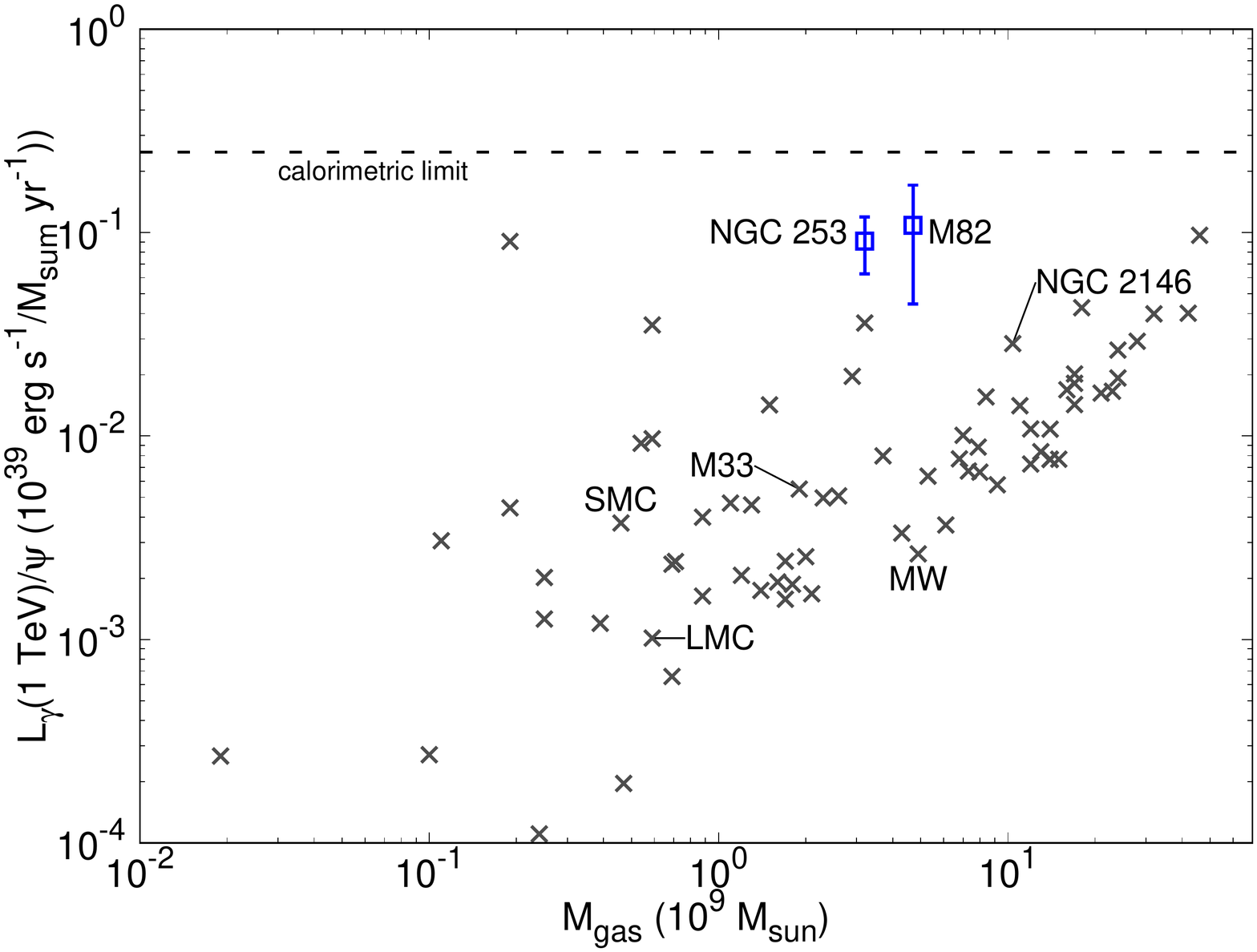}
\end{center}
\end{minipage}
\caption{Correlation between $L_\gamma({\rm 1\,TeV})/\psi$ and gas
  surface density
  (left), or gas mass (right), of nearby galaxies predicted by our
  model assuming $\Gamma_\mathrm{inj}=2.2$.
  The dashed line is the calorimetric limit,
  $L_\gamma(1\,{\rm TeV})/{\rm SFR}\sim 2.5\times10^{38}\,{\rm erg\,s^{-1}/M_\odot\,yr^{-1}}$ (see Section~\ref{section:sample}).
  Blue points are galaxies detected by 
  TeV gamma-ray telescopes based on the observations by \citet{H.E.S.S.Collaboration2018} and \citet{VERITASCollaboration2009}.
}
\label{fig:TeV-relation}
\end{figure*}

To examine this, we made a plot of $L_\gamma(1\,{\rm TeV})/\psi$ versus
${ \Sigma_{\rm gas}=M_{\rm gas}/(\pi R_{\rm eff}^2)}$ (surface gas density) in
Figure~\ref{fig:TeV-relation} (left panel).  If $\Sigma_{\rm gas}$ is
sufficiently high, cosmic-ray collision efficiency would be high and
gamma-ray emission would be close to the calorimetric limit (dashed line), and hence
$L_\gamma(1\,{\rm TeV})/\psi$ will become constant asymptotically. This trend
is indeed seen in the figure, and M82, NGC 1482 and DDO 165 are located at
the highest $\Sigma_{\rm gas}$ region.  The spectrum of NGC 1482
predicted by our model is shown in Figure~\ref{fig:spectrum_examples}
(bottom left). Comparing to the CTA sensitivity, detection may not be
easy but not impossible depending on model parameters and
uncertainties.  If detected, we predict that this galaxy would have a
particularly hard spectrum compared with other SFGs.  On the other
hand, the predicted flux of DDO 165 is far below the CTA sensitivity
limit (see Figure~\ref{fig:flux-flux}).

\begin{figure*}
\begin{minipage}{0.45\hsize}
\begin{center}
\includegraphics[width=\columnwidth]{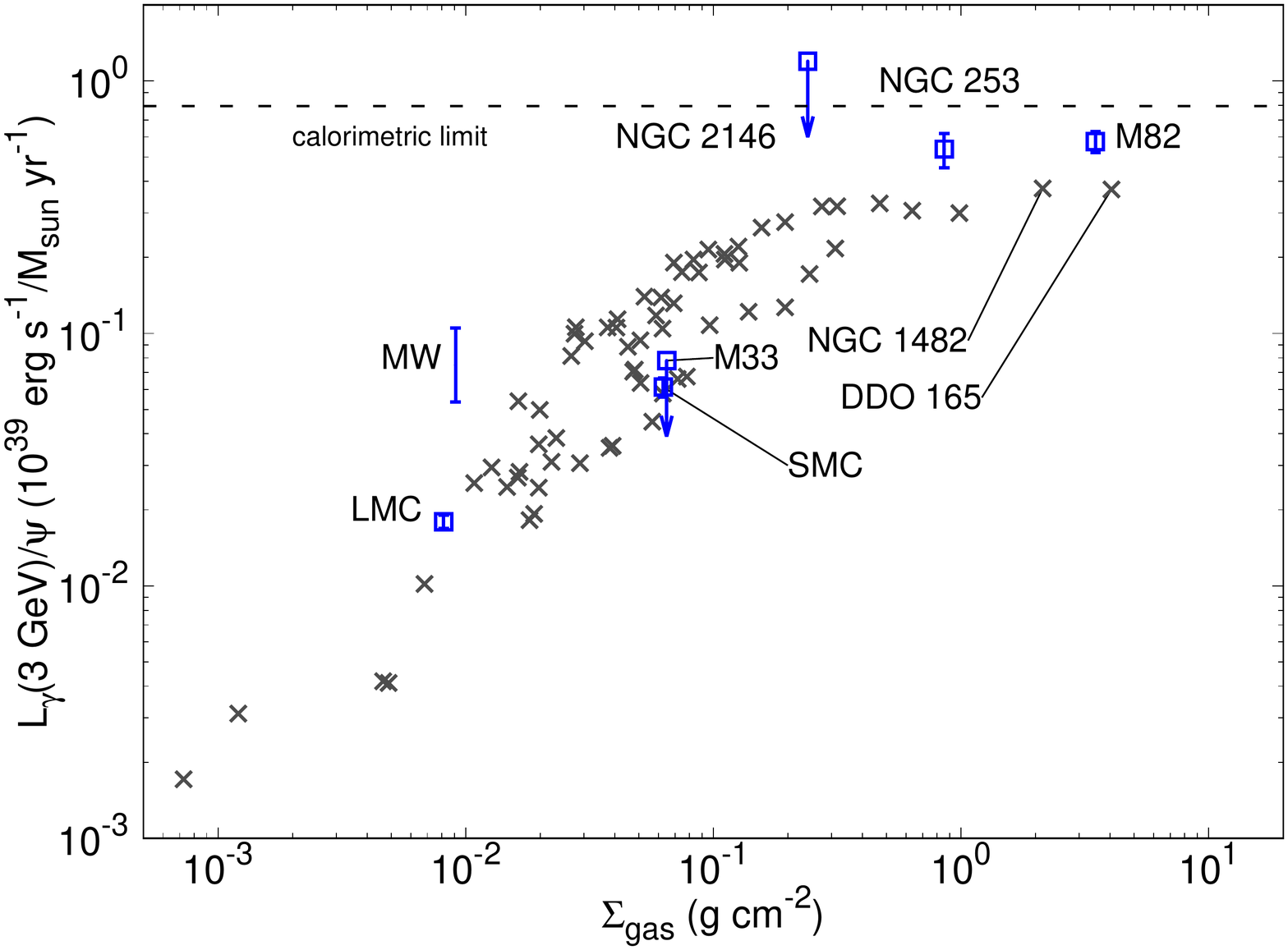}
\end{center}
\end{minipage}
\begin{minipage}{0.45\hsize}
\begin{center}
\includegraphics[width=\columnwidth]{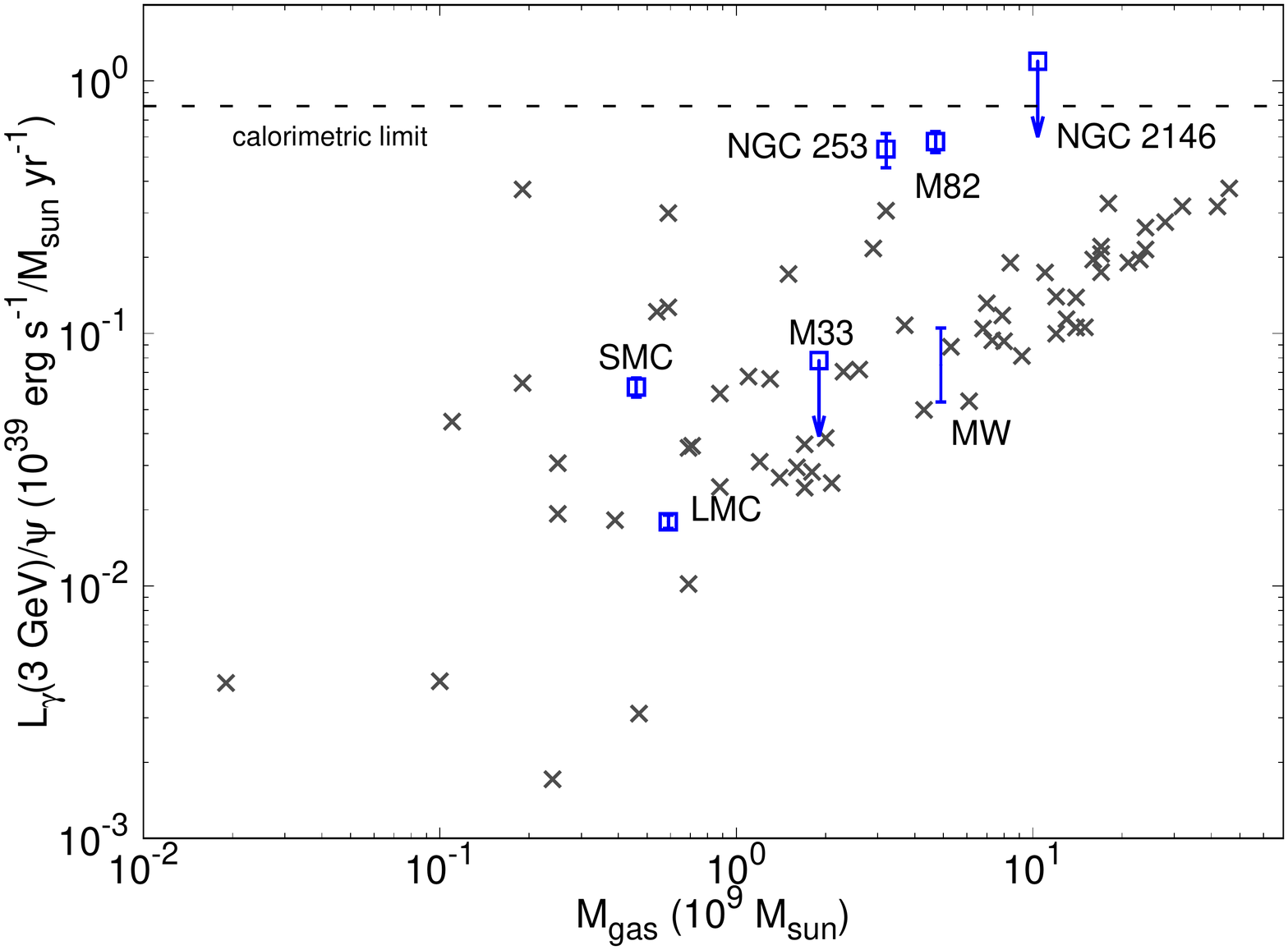}
\end{center}
\end{minipage}
\caption{The same as Fig. \ref{fig:TeV-relation}, but for luminosities at 3 GeV.  
The dashed line is the calorimetric limit,
$L_\gamma(3\,{\rm GeV})/{\rm SFR}\sim 7.9\times10^{38}\,{\rm erg\,s^{-1}/M_\odot\,yr^{-1}}$
(the same assumption as the TeV calorimetric limit; see Section~\ref{section:sample}).
Blue points are galaxies detected by {\it Fermi}-LAT based on the observations by \citet{Ajello2020}.  The
  data point for MW is based on \citet{strong2010}, showing the
  range between the highest and lowest luminosities of their models.
}
\label{fig:GeV-relation}
\end{figure*}

Gamma-ray luminosities of SFGs are often compared
with $\psi M_{\rm gas}$, with an expectation of $L_\gamma \propto \psi
M_{\rm gas}$. Here, the amount of gas mass is considered as an
indicator of cosmic-ray collision efficiency. To examine this, a
correlation plot between $L(1\,{\rm TeV})/\psi$ and $M_{\rm gas}$ is also
shown in the right panel of Figure~\ref{fig:TeV-relation}. The data
points show a positive correlation similar to the case of the
$L(1\,{\rm TeV})/\psi$-$\Sigma_{\rm gas}$ correlation, but as expected, the scatter is
much larger because the information of galaxy sizes is not used.  This
indicates a limitation of using $\psi M_{\rm gas}$ as an indicator of
gamma-ray luminosity. The same correlations between $L(3\,{\rm GeV})/\psi$
versus $\Sigma_{\rm gas}$ or $M_{\rm gas}$
are shown in Figure~\ref{fig:GeV-relation}.  In both the
energy bands, $\Sigma_{\rm gas}$ is a good indicator of $L_\gamma/\psi$. Compared
with the relation in TeV, the relation in GeV becomes flat at $\Sigma_{\rm gas}
\gtrsim 1 \ \rm g \ cm^{-2}$. 
This is expected because GeV emission would reach the calorimetric
limit at lower $\Sigma_{\rm gas}$.

\section{Discussion \& Conclusions}
\label{section:conclusions}
In this work, we calculated expected fluxes and spectra of TeV gamma
rays from nearby SFGs, mainly included in the catalog of KINGFISH,
based on the four physical quantities of galaxies: SFR, gas mass,
stellar mass, and effective radius.  Calculations were done using the
S18 model, which considers energy-dependent propagation and escape of
cosmic rays.  These model predictions are more reliable than simpler
estimates based only on SFRs. Predicted gamma-ray fluxes at 1 TeV
are tabulated in Table \ref{tab:calculated-flux}, for
several values of the assumed spectral index $\Gamma_{\rm inj}$ of
cosmic-ray protons injected into ISM. Our recommendation
is $\Gamma_{\rm inj} = 2.2$ because this value is close to
those (2.1--2.3)
favored for MW, NGC 253, and M82, though
other galaxies may have a significantly different value.

Then we examined the detectability of these galaxies by CTA.  There is
a good chance of detection for NGC 5236, M33, NGC 6946, and IC 342.
Our model predicts that power-law spectral indices of gamma-ray
emission in the CTA band are typically 0.20--0.30 softer than those of
injected protons. However, some galaxies having particularly high
surface gas densities, such as M82, NGC 1482 and DDO 165, have harder
spectra with a photon index softer only by at most 0.14 than protons. It
is interesting that NGC 1482 may also be detected by CTA, and we
predict that this galaxy would have a particularly hard TeV spectrum
than other SFGs if the proton injection spectrum is universal.  TeV
gamma-ray luminosities of most galaxies in the sample predicted by our
model are lower than the simple estimates based only on SFR and
assuming the same $L_\gamma({\rm 1\,TeV})/\psi$ ratio as NGC 253. The
$L_\gamma({\rm 1\,TeV})/\psi$ ratio increases with surface gas
densities of galaxies. These results indicate that most nearby
galaxies are not in the calorimetric limit and cosmic-ray escape from
galaxies is important.

Recently detection of GeV gamma rays from NGC 2403 by {\it Fermi}-LAT was
reported \citep{Xi2020b}, and the gamma-ray flux seems variable.  Then
it was proposed that this gamma-ray emission is coming from a
coincident supernova 2004dj. However, flux variability is
statistically marginal, and a possibility of gamma-ray emission from
other origins remains. Therefore we show here the gamma-ray spectrum
of NGC 2403 predicted by our model in
Figure~\ref{fig:spectrum_examples}. The observed flux is about an
order of magnitude higher than the expected flux, and it is unlikely
that star-formation activity and cosmic-ray interaction explain the
observed flux. Therefore the observed GeV flux should be explained
by the supernova or any other gamma-ray emission processes.

Throughout, we have focused on gamma-ray emission produced by hadronic
processes. This assumption is believed to be secure in the GeV
range. However, recent studies have found that leptonic emission
powered by pulsars may significantly contribute to, or even dominate,
the diffuse Galactic gamma-ray emission in the TeV regime
\citep{2018PhRvL.120l1101L, 2019PhRvD.100d3016S}. If future CTA
observations detect SFGs with fluxes larger than predicted by our
hadronic model, it may suggest a contribution from leptons, offering
support for the importance of pulsar emission in the TeV sky.

\section*{Acknowledgements}
This work was conducted in the context of the CTA Consortium.
This research has made use of the CTA instrument response functions provided by the CTA Consortium and Observatory, see \url{http://www.cta-observatory.org/science/cta-performance/} (version prod3b-v1) for more details.
TT was supported by the JSPS/MEXT KAKENHI Grant Numbers 18K03692 and
17H06362.  T.S. is supported by a Research Fellowship of Japan Society
for the Promotion of Science (JSPS) and by JSPS KAKENHI Grant No.\ JP
18J20943.

\section*{Data Availability}
The data underlying this article will be shared on reasonable request to the corresponding author.
The full lists for Table~\ref{tab:calc-parameter} and Table~\ref{tab:calculated-flux} are available in their online supplementary materials.n%




\bibliographystyle{mnras}
\bibliography{reference} 

\begin{thebibliography}{}
\makeatletter
\relax
\def\mn@urlcharsother{\let\do\@makeother \do\$\do\&\do\#\do\^\do\_\do\%\do\~}
\def\mn@doi{\begingroup\mn@urlcharsother \@ifnextchar [ {\mn@doi@}
  {\mn@doi@[]}}
\def\mn@doi@[#1]#2{\def\@tempa{#1}\ifx\@tempa\@empty \href
  {http://dx.doi.org/#2} {doi:#2}\else \href {http://dx.doi.org/#2} {#1}\fi
  \endgroup}
\def\mn@eprint#1#2{\mn@eprint@#1:#2::\@nil}
\def\mn@eprint@arXiv#1{\href {http://arxiv.org/abs/#1} {{\tt arXiv:#1}}}
\def\mn@eprint@dblp#1{\href {http://dblp.uni-trier.de/rec/bibtex/#1.xml}
  {dblp:#1}}
\def\mn@eprint@#1:#2:#3:#4\@nil{\def\@tempa {#1}\def\@tempb {#2}\def\@tempc
  {#3}\ifx \@tempc \@empty \let \@tempc \@tempb \let \@tempb \@tempa \fi \ifx
  \@tempb \@empty \def\@tempb {arXiv}\fi \@ifundefined
  {mn@eprint@\@tempb}{\@tempb:\@tempc}{\expandafter \expandafter \csname
  mn@eprint@\@tempb\endcsname \expandafter{\@tempc}}}

\bibitem[\protect\citeauthoryear{{Abdo} et~al.,}{{Abdo}
  et~al.}{2009}]{Abdo2009}
{Abdo} A.~A.,  et~al., 2009, \mn@doi [\prl] {10.1103/PhysRevLett.103.251101},
  \href {https://ui.adsabs.harvard.edu/abs/2009PhRvL.103y1101A} {103, 251101}

\bibitem[\protect\citeauthoryear{{Abdo} et~al.,}{{Abdo}
  et~al.}{2010a}]{Abdo2010a}
{Abdo} A.~A.,  et~al., 2010a, \mn@doi [\aap] {10.1051/0004-6361/200913474},
  \href {https://ui.adsabs.harvard.edu/abs/2010A&A...512A...7A} {512, A7}

\bibitem[\protect\citeauthoryear{{Abdo} et~al.,}{{Abdo}
  et~al.}{2010b}]{Abdo2010b}
{Abdo} A.~A.,  et~al., 2010b, \mn@doi [\aap] {10.1051/0004-6361/201015759},
  \href {https://ui.adsabs.harvard.edu/abs/2010A&A...523L...2A} {523, L2}

\bibitem[\protect\citeauthoryear{{Abdo} et~al.,}{{Abdo}
  et~al.}{2010c}]{Abdo2010c}
{Abdo} A.~A.,  et~al., 2010c, \mn@doi [\aap] {10.1051/0004-6361/201014855},
  \href {https://ui.adsabs.harvard.edu/abs/2010A&A...523A..46A} {523, A46}

\bibitem[\protect\citeauthoryear{{Abdo} et~al.,}{{Abdo}
  et~al.}{2010d}]{Abdo2010d}
{Abdo} A.~A.,  et~al., 2010d, \mn@doi [\apjl] {10.1088/2041-8205/709/2/L152},
  \href {https://ui.adsabs.harvard.edu/abs/2010ApJ...709L.152A} {709, L152}

\bibitem[\protect\citeauthoryear{{Acero} et~al.,}{{Acero}
  et~al.}{2009}]{Acero2009}
{Acero} F.,  et~al., 2009, \mn@doi [Science] {10.1126/science.1178826}, \href
  {https://ui.adsabs.harvard.edu/abs/2009Sci...326.1080A} {326, 1080}

\bibitem[\protect\citeauthoryear{{Ackermann} et~al.,}{{Ackermann}
  et~al.}{2012a}]{Ackermann2012b}
{Ackermann} M.,  et~al., 2012a, \mn@doi [\apj] {10.1088/0004-637X/750/1/3},
  \href {https://ui.adsabs.harvard.edu/abs/2012ApJ...750....3A} {750, 3}

\bibitem[\protect\citeauthoryear{{Ackermann} et~al.,}{{Ackermann}
  et~al.}{2012b}]{Ackermann2012}
{Ackermann} M.,  et~al., 2012b, \mn@doi [\apj] {10.1088/0004-637X/755/2/164},
  \href {https://ui.adsabs.harvard.edu/abs/2012ApJ...755..164A} {755, 164}

\bibitem[\protect\citeauthoryear{{Ackermann} et~al.,}{{Ackermann}
  et~al.}{2017}]{Ackermann2017}
{Ackermann} M.,  et~al., 2017, \mn@doi [\apj] {10.3847/1538-4357/aa5c3d}, \href
  {https://ui.adsabs.harvard.edu/abs/2017ApJ...836..208A} {836, 208}

\bibitem[\protect\citeauthoryear{{Adams} \& {Weedman}}{{Adams} \&
  {Weedman}}{1975}]{Adams1975}
{Adams} T.~F.,  {Weedman} D.~W.,  1975, \mn@doi [\apj] {10.1086/153659}, \href
  {https://ui.adsabs.harvard.edu/abs/1975ApJ...199...19A} {199, 19}

\bibitem[\protect\citeauthoryear{{Ajello}, {Di Mauro}, {Paliya}  \&
  {Garrappa}}{{Ajello} et~al.}{2020}]{Ajello2020}
{Ajello} M.,  {Di Mauro} M.,  {Paliya} V.~S.,   {Garrappa} S.,  2020, \mn@doi
  [\apj] {10.3847/1538-4357/ab86a6}, \href
  {https://ui.adsabs.harvard.edu/abs/2020ApJ...894...88A} {894, 88}

\bibitem[\protect\citeauthoryear{{Allison}, {Sadler}  \& {Meekin}}{{Allison}
  et~al.}{2014}]{Allison2014}
{Allison} J.~R.,  {Sadler} E.~M.,   {Meekin} A.~M.,  2014, \mn@doi [\mnras]
  {10.1093/mnras/stu289}, \href
  {https://ui.adsabs.harvard.edu/abs/2014MNRAS.440..696A} {440, 696}

\bibitem[\protect\citeauthoryear{{Aloisio} \& {Berezinsky}}{{Aloisio} \&
  {Berezinsky}}{2004}]{Aloisio2004}
{Aloisio} R.,  {Berezinsky} V.,  2004, \mn@doi [\apj] {10.1086/421869}, \href
  {https://ui.adsabs.harvard.edu/abs/2004ApJ...612..900A} {612, 900}

\bibitem[\protect\citeauthoryear{{Ambrosone}, {Chianese}, {Fiorillo},
  {Marinelli}, {Miele}  \& {Pisanti}}{{Ambrosone} et~al.}{2021}]{Ambrosone2021}
{Ambrosone} A.,  {Chianese} M.,  {Fiorillo} D. F.~G.,  {Marinelli} A.,  {Miele}
  G.,   {Pisanti} O.,  2021, \mn@doi [\mnras] {10.1093/mnras/stab659}, \href
  {https://ui.adsabs.harvard.edu/abs/2021MNRAS.tmp..765A} {}

\bibitem[\protect\citeauthoryear{{Armus}, {Heckman}  \& {Miley}}{{Armus}
  et~al.}{1990}]{Armus1990}
{Armus} L.,  {Heckman} T.~M.,   {Miley} G.~K.,  1990, \mn@doi [\apj]
  {10.1086/169431}, \href
  {https://ui.adsabs.harvard.edu/abs/1990ApJ...364..471A} {364, 471}

\bibitem[\protect\citeauthoryear{{Beck}}{{Beck}}{2008}]{Beck2008}
{Beck} R.,  2008, in {Aharonian} F.~A.,  {Hofmann} W.,   {Rieger} F.,  eds,
  American Institute of Physics Conference Series Vol. 1085, American Institute
  of Physics Conference Series. pp 83--96 (\mn@eprint {arXiv} {0810.2923}),
  \mn@doi{10.1063/1.3076806}

\bibitem[\protect\citeauthoryear{{Bland-Hawthorn} \&
  {Gerhard}}{{Bland-Hawthorn} \& {Gerhard}}{2016}]{BlandHawthorn2016}
{Bland-Hawthorn} J.,  {Gerhard} O.,  2016, \mn@doi [\araa]
  {10.1146/annurev-astro-081915-023441}, \href
  {https://ui.adsabs.harvard.edu/abs/2016ARA&A..54..529B} {54, 529}

\bibitem[\protect\citeauthoryear{{Braun}, {Thilker}, {Walterbos}  \&
  {Corbelli}}{{Braun} et~al.}{2009}]{Braun2009}
{Braun} R.,  {Thilker} D.~A.,  {Walterbos} R.~A.~M.,   {Corbelli} E.,  2009,
  \mn@doi [\apj] {10.1088/0004-637X/695/2/937}, \href
  {https://ui.adsabs.harvard.edu/abs/2009ApJ...695..937B} {695, 937}

\bibitem[\protect\citeauthoryear{{Brown} et~al.,}{{Brown}
  et~al.}{2014}]{Brown2014}
{Brown} M. J.~I.,  et~al., 2014, \mn@doi [\apjs] {10.1088/0067-0049/212/2/18},
  \href {https://ui.adsabs.harvard.edu/abs/2014ApJS..212...18B} {212, 18}

\bibitem[\protect\citeauthoryear{{Chan}, {Kere{\v{s}}}, {Hopkins}, {Quataert},
  {Su}, {Hayward}  \& {Faucher-Gigu{\`e}re}}{{Chan} et~al.}{2019}]{Chan2019}
{Chan} T.~K.,  {Kere{\v{s}}} D.,  {Hopkins} P.~F.,  {Quataert} E.,  {Su} K.~Y.,
   {Hayward} C.~C.,   {Faucher-Gigu{\`e}re} C.~A.,  2019, \mn@doi [\mnras]
  {10.1093/mnras/stz1895}, \href
  {https://ui.adsabs.harvard.edu/abs/2019MNRAS.488.3716C} {488, 3716}

\bibitem[\protect\citeauthoryear{{Cherenkov Telescope Array Consortium}
  et~al.,}{{Cherenkov Telescope Array Consortium}
  et~al.}{2019}]{CherenkovTelescopeArrayConsortium2019}
{Cherenkov Telescope Array Consortium} et~al., 2019, {Science with the
  Cherenkov Telescope Array}, \mn@doi{10.1142/10986.
}

\bibitem[\protect\citeauthoryear{{Chynoweth}, {Langston}, {Yun}, {Lockman},
  {Rubin}  \& {Scoles}}{{Chynoweth} et~al.}{2008}]{Chynoweth2008}
{Chynoweth} K.~M.,  {Langston} G.~I.,  {Yun} M.~S.,  {Lockman} F.~J.,  {Rubin}
  K.~H.~R.,   {Scoles} S.~A.,  2008, \mn@doi [\aj]
  {10.1088/0004-6256/135/6/1983}, \href
  {https://ui.adsabs.harvard.edu/abs/2008AJ....135.1983C} {135, 1983}

\bibitem[\protect\citeauthoryear{{Dale} \& {Helou}}{{Dale} \&
  {Helou}}{2002}]{Dale2002}
{Dale} D.~A.,  {Helou} G.,  2002, \mn@doi [\apj] {10.1086/341632}, \href
  {https://ui.adsabs.harvard.edu/abs/2002ApJ...576..159D} {576, 159}

\bibitem[\protect\citeauthoryear{{Dale} et~al.,}{{Dale}
  et~al.}{2009}]{Dale2009}
{Dale} D.~A.,  et~al., 2009, \mn@doi [\apj] {10.1088/0004-637X/703/1/517},
  \href {https://ui.adsabs.harvard.edu/abs/2009ApJ...703..517D} {703, 517}

\bibitem[\protect\citeauthoryear{{Diehl} et~al.,}{{Diehl}
  et~al.}{2006}]{Diehl2006}
{Diehl} R.,  et~al., 2006, \mn@doi [\nat] {10.1038/nature04364}, \href
  {https://ui.adsabs.harvard.edu/abs/2006Natur.439...45D} {439, 45}

\bibitem[\protect\citeauthoryear{{Domingo-Santamar{\'\i}a} \&
  {Torres}}{{Domingo-Santamar{\'\i}a} \&
  {Torres}}{2005}]{DomingoSantamaria2005}
{Domingo-Santamar{\'\i}a} E.,  {Torres} D.~F.,  2005, \mn@doi [\aap]
  {10.1051/0004-6361:20053613}, \href
  {https://ui.adsabs.harvard.edu/abs/2005A&A...444..403D} {444, 403}

\bibitem[\protect\citeauthoryear{{Eichmann} \& {Becker Tjus}}{{Eichmann} \&
  {Becker Tjus}}{2016}]{Eichmann2016}
{Eichmann} B.,  {Becker Tjus} J.,  2016, \mn@doi [\apj]
  {10.3847/0004-637X/821/2/87}, \href
  {https://ui.adsabs.harvard.edu/abs/2016ApJ...821...87E} {821, 87}

\bibitem[\protect\citeauthoryear{{Eskew}, {Zaritsky}  \& {Meidt}}{{Eskew}
  et~al.}{2012}]{Eskew2012}
{Eskew} M.,  {Zaritsky} D.,   {Meidt} S.,  2012, \mn@doi [\aj]
  {10.1088/0004-6256/143/6/139}, \href
  {https://ui.adsabs.harvard.edu/abs/2012AJ....143..139E} {143, 139}

\bibitem[\protect\citeauthoryear{{Foley} et~al.,}{{Foley}
  et~al.}{2014}]{Foley2014}
{Foley} R.~J.,  et~al., 2014, \mn@doi [\mnras] {10.1093/mnras/stu1378}, \href
  {https://ui.adsabs.harvard.edu/abs/2014MNRAS.443.2887F} {443, 2887}

\bibitem[\protect\citeauthoryear{{Galametz}, {Madden}, {Galliano}, {Hony},
  {Bendo}  \& {Sauvage}}{{Galametz} et~al.}{2011}]{Galametz2011}
{Galametz} M.,  {Madden} S.~C.,  {Galliano} F.,  {Hony} S.,  {Bendo} G.~J.,
  {Sauvage} M.,  2011, \mn@doi [\aap] {10.1051/0004-6361/201014904}, \href
  {https://ui.adsabs.harvard.edu/abs/2011A&A...532A..56G} {532, A56}

\bibitem[\protect\citeauthoryear{{Gonidakis}, {Livanou}, {Kontizas}, {Klein},
  {Kontizas}, {Belcheva}, {Tsalmantza}  \& {Karampelas}}{{Gonidakis}
  et~al.}{2009}]{Gonidakis2009}
{Gonidakis} I.,  {Livanou} E.,  {Kontizas} E.,  {Klein} U.,  {Kontizas} M.,
  {Belcheva} M.,  {Tsalmantza} P.,   {Karampelas} A.,  2009, \mn@doi [\aap]
  {10.1051/0004-6361/200809828}, \href
  {https://ui.adsabs.harvard.edu/abs/2009A&A...496..375G} {496, 375}

\bibitem[\protect\citeauthoryear{{Gratier} et~al.,}{{Gratier}
  et~al.}{2010}]{Gratier2010}
{Gratier} P.,  et~al., 2010, \mn@doi [\aap] {10.1051/0004-6361/201014441},
  \href {https://ui.adsabs.harvard.edu/abs/2010A&A...522A...3G} {522, A3}

\bibitem[\protect\citeauthoryear{{H.~E.~S.~S. Collaboration}
  et~al.,}{{H.~E.~S.~S. Collaboration}
  et~al.}{2018}]{H.E.S.S.Collaboration2018}
{H.~E.~S.~S. Collaboration} et~al., 2018, \mn@doi [\aap]
  {10.1051/0004-6361/201833202}, \href
  {https://ui.adsabs.harvard.edu/abs/2018A&A...617A..73H} {617, A73}

\bibitem[\protect\citeauthoryear{{Haynes} et~al.,}{{Haynes}
  et~al.}{2018}]{Haynes2018}
{Haynes} M.~P.,  et~al., 2018, \mn@doi [\apj] {10.3847/1538-4357/aac956}, \href
  {https://ui.adsabs.harvard.edu/abs/2018ApJ...861...49H} {861, 49}

\bibitem[\protect\citeauthoryear{{Heyer}, {Corbelli}, {Schneider}  \&
  {Young}}{{Heyer} et~al.}{2004}]{Heyer2004}
{Heyer} M.~H.,  {Corbelli} E.,  {Schneider} S.~E.,   {Young} J.~S.,  2004,
  \mn@doi [\apj] {10.1086/381196}, \href
  {https://ui.adsabs.harvard.edu/abs/2004ApJ...602..723H} {602, 723}

\bibitem[\protect\citeauthoryear{{Holler} et~al.,}{{Holler}
  et~al.}{2015}]{Holler2015}
{Holler} M.,  et~al., 2015, arXiv e-prints, \href
  {https://ui.adsabs.harvard.edu/abs/2015arXiv150902902H} {p. arXiv:1509.02902}

\bibitem[\protect\citeauthoryear{{Howell} et~al.,}{{Howell}
  et~al.}{2007}]{Howell2007}
{Howell} J.~H.,  et~al., 2007, \mn@doi [\aj] {10.1086/521821}, \href
  {https://ui.adsabs.harvard.edu/abs/2007AJ....134.2086H} {134, 2086}

\bibitem[\protect\citeauthoryear{{Inoue}}{{Inoue}}{2011}]{Inoue2011}
{Inoue} Y.,  2011, \mn@doi [\apj] {10.1088/0004-637X/728/1/11}, \href
  {https://ui.adsabs.harvard.edu/abs/2011ApJ...728...11I} {728, 11}

\bibitem[\protect\citeauthoryear{{Inoue}, {Inoue}, {Kobayashi}, {Makiya},
  {Niino}  \& {Totani}}{{Inoue} et~al.}{2013}]{Inoue2013}
{Inoue} Y.,  {Inoue} S.,  {Kobayashi} M. A.~R.,  {Makiya} R.,  {Niino} Y.,
  {Totani} T.,  2013, \mn@doi [\apj] {10.1088/0004-637X/768/2/197}, \href
  {https://ui.adsabs.harvard.edu/abs/2013ApJ...768..197I} {768, 197}

\bibitem[\protect\citeauthoryear{{Israel}}{{Israel}}{1997}]{Israel1997}
{Israel} F.~P.,  1997, \aap, \href
  {https://ui.adsabs.harvard.edu/abs/1997A&A...328..471I} {328, 471}

\bibitem[\protect\citeauthoryear{{Jarrett}, {Chester}, {Cutri}, {Schneider}  \&
  {Huchra}}{{Jarrett} et~al.}{2003}]{Jarrett2003}
{Jarrett} T.~H.,  {Chester} T.,  {Cutri} R.,  {Schneider} S.~E.,   {Huchra}
  J.~P.,  2003, \mn@doi [\aj] {10.1086/345794}, \href
  {https://ui.adsabs.harvard.edu/abs/2003AJ....125..525J} {125, 525}

\bibitem[\protect\citeauthoryear{{Kelner}, {Aharonian}  \& {Bugayov}}{{Kelner}
  et~al.}{2006}]{Kelner2006}
{Kelner} S.~R.,  {Aharonian} F.~A.,   {Bugayov} V.~V.,  2006, \mn@doi [\prd]
  {10.1103/PhysRevD.74.034018}, \href
  {https://ui.adsabs.harvard.edu/abs/2006PhRvD..74c4018K} {74, 034018}

\bibitem[\protect\citeauthoryear{{Kennicutt}, {Lee}, {Funes}, {J.}, {Sakai}  \&
  {Akiyama}}{{Kennicutt} et~al.}{2008}]{Kennicutt2008}
{Kennicutt} Robert~C. J.,  {Lee} J.~C.,  {Funes} J.~G.,  {J.} S.,  {Sakai} S.,
   {Akiyama} S.,  2008, \mn@doi [\apjs] {10.1086/590058}, \href
  {https://ui.adsabs.harvard.edu/abs/2008ApJS..178..247K} {178, 247}

\bibitem[\protect\citeauthoryear{{Kennicutt} Robert~C. et~al.,}{{Kennicutt}
  et~al.}{2009}]{Kennicutt2009}
{Kennicutt} Robert~C. J.,  et~al., 2009, \mn@doi [\apj]
  {10.1088/0004-637X/703/2/1672}, \href
  {https://ui.adsabs.harvard.edu/abs/2009ApJ...703.1672K} {703, 1672}

\bibitem[\protect\citeauthoryear{{Kennicutt} et~al.,}{{Kennicutt}
  et~al.}{2011}]{Kennicutt2011}
{Kennicutt} R.~C.,  et~al., 2011, \mn@doi [\pasp] {10.1086/663818}, \href
  {https://ui.adsabs.harvard.edu/abs/2011PASP..123.1347K} {123, 1347}

\bibitem[\protect\citeauthoryear{{Knudsen}, {Walter}, {Weiss}, {Bolatto},
  {Riechers}  \& {Menten}}{{Knudsen} et~al.}{2007}]{Knudsen2007}
{Knudsen} K.~K.,  {Walter} F.,  {Weiss} A.,  {Bolatto} A.,  {Riechers} D.~A.,
  {Menten} K.,  2007, \mn@doi [\apj] {10.1086/519761}, \href
  {https://ui.adsabs.harvard.edu/abs/2007ApJ...666..156K} {666, 156}

\bibitem[\protect\citeauthoryear{{Krumholz}, {Crocker}, {Xu}, {Lazarian},
  {Rosevear}  \& {Bedwell-Wilson}}{{Krumholz} et~al.}{2020}]{Krumholz2020}
{Krumholz} M.~R.,  {Crocker} R.~M.,  {Xu} S.,  {Lazarian} A.,  {Rosevear}
  M.~T.,   {Bedwell-Wilson} J.,  2020, \mn@doi [\mnras]
  {10.1093/mnras/staa493}, \href
  {https://ui.adsabs.harvard.edu/abs/2020MNRAS.493.2817K} {493, 2817}

\bibitem[\protect\citeauthoryear{{Lacki} \& {Thompson}}{{Lacki} \&
  {Thompson}}{2013}]{Lacki2013}
{Lacki} B.~C.,  {Thompson} T.~A.,  2013, \mn@doi [\apj]
  {10.1088/0004-637X/762/1/29}, \href
  {https://ui.adsabs.harvard.edu/abs/2013ApJ...762...29L} {762, 29}

\bibitem[\protect\citeauthoryear{{Lacki}, {Thompson}  \& {Quataert}}{{Lacki}
  et~al.}{2010}]{Lacki2010}
{Lacki} B.~C.,  {Thompson} T.~A.,   {Quataert} E.,  2010, \mn@doi [\apj]
  {10.1088/0004-637X/717/1/1}, \href
  {https://ui.adsabs.harvard.edu/abs/2010ApJ...717....1L} {717, 1}

\bibitem[\protect\citeauthoryear{{Lacki}, {Thompson}, {Quataert}, {Loeb}  \&
  {Waxman}}{{Lacki} et~al.}{2011}]{Lacki2011}
{Lacki} B.~C.,  {Thompson} T.~A.,  {Quataert} E.,  {Loeb} A.,   {Waxman} E.,
  2011, \mn@doi [\apj] {10.1088/0004-637X/734/2/107}, \href
  {https://ui.adsabs.harvard.edu/abs/2011ApJ...734..107L} {734, 107}

\bibitem[\protect\citeauthoryear{{Leroy}, {Walter}, {Brinks}, {Bigiel}, {de
  Blok}, {Madore}  \& {Thornley}}{{Leroy} et~al.}{2008}]{Leroy2008}
{Leroy} A.~K.,  {Walter} F.,  {Brinks} E.,  {Bigiel} F.,  {de Blok} W.~J.~G.,
  {Madore} B.,   {Thornley} M.~D.,  2008, \mn@doi [\aj]
  {10.1088/0004-6256/136/6/2782}, \href
  {https://ui.adsabs.harvard.edu/abs/2008AJ....136.2782L} {136, 2782}

\bibitem[\protect\citeauthoryear{{Linden} \& {Buckman}}{{Linden} \&
  {Buckman}}{2018}]{2018PhRvL.120l1101L}
{Linden} T.,  {Buckman} B.~J.,  2018, \mn@doi [\prl]
  {10.1103/PhysRevLett.120.121101}, \href
  {https://ui.adsabs.harvard.edu/abs/2018PhRvL.120l1101L} {120, 121101}

\bibitem[\protect\citeauthoryear{{Lopez}, {Auchettl}, {Linden}, {Bolatto},
  {Thompson}  \& {Ramirez-Ruiz}}{{Lopez} et~al.}{2018}]{Lopez2018}
{Lopez} L.~A.,  {Auchettl} K.,  {Linden} T.,  {Bolatto} A.~D.,  {Thompson}
  T.~A.,   {Ramirez-Ruiz} E.,  2018, \mn@doi [\apj] {10.3847/1538-4357/aae0f8},
  \href {https://ui.adsabs.harvard.edu/abs/2018ApJ...867...44L} {867, 44}

\bibitem[\protect\citeauthoryear{{Madden} et~al.,}{{Madden}
  et~al.}{2013}]{Madden2013}
{Madden} S.~C.,  et~al., 2013, \mn@doi [\pasp] {10.1086/671138}, \href
  {https://ui.adsabs.harvard.edu/abs/2013PASP..125..600M} {125, 600}

\bibitem[\protect\citeauthoryear{{Makiya}, {Totani}  \& {Kobayashi}}{{Makiya}
  et~al.}{2011}]{Makiya2011}
{Makiya} R.,  {Totani} T.,   {Kobayashi} M. A.~R.,  2011, \mn@doi [\apj]
  {10.1088/0004-637X/728/2/158}, \href
  {https://ui.adsabs.harvard.edu/abs/2011ApJ...728..158M} {728, 158}

\bibitem[\protect\citeauthoryear{{Marble} et~al.,}{{Marble}
  et~al.}{2010}]{Marble2010}
{Marble} A.~R.,  et~al., 2010, \mn@doi [\apj] {10.1088/0004-637X/715/1/506},
  \href {https://ui.adsabs.harvard.edu/abs/2010ApJ...715..506M} {715, 506}

\bibitem[\protect\citeauthoryear{{Martin}}{{Martin}}{2014}]{Martin2014}
{Martin} P.,  2014, \mn@doi [\aap] {10.1051/0004-6361/201323329}, \href
  {https://ui.adsabs.harvard.edu/abs/2014A&A...564A..61M} {564, A61}

\bibitem[\protect\citeauthoryear{{Mo}, {van den Bosch}  \& {White}}{{Mo}
  et~al.}{2010}]{Mo2010}
{Mo} H.,  {van den Bosch} F.~C.,   {White} S.,  2010, {Galaxy Formation and
  Evolution}

\bibitem[\protect\citeauthoryear{{Montalto}, {Seitz}, {Riffeser}, {Hopp}, {Lee}
   \& {Sch{\"o}nrich}}{{Montalto} et~al.}{2009}]{Montalto2009}
{Montalto} M.,  {Seitz} S.,  {Riffeser} A.,  {Hopp} U.,  {Lee} C.~H.,
  {Sch{\"o}nrich} R.,  2009, \mn@doi [\aap] {10.1051/0004-6361/200912179},
  \href {https://ui.adsabs.harvard.edu/abs/2009A&A...507..283M} {507, 283}

\bibitem[\protect\citeauthoryear{{Moustakas}, {Kennicutt}, {Tremonti}, {Dale},
  {Smith}  \& {Calzetti}}{{Moustakas} et~al.}{2010}]{Moustakas2010}
{Moustakas} J.,  {Kennicutt} Robert~C. J.,  {Tremonti} C.~A.,  {Dale} D.~A.,
  {Smith} J.-D.~T.,   {Calzetti} D.,  2010, \mn@doi [\apjs]
  {10.1088/0067-0049/190/2/233}, \href
  {https://ui.adsabs.harvard.edu/abs/2010ApJS..190..233M} {190, 233}

\bibitem[\protect\citeauthoryear{{Nieten}, {Neininger}, {Gu{\'e}lin},
  {Ungerechts}, {Lucas}, {Berkhuijsen}, {Beck}  \& {Wielebinski}}{{Nieten}
  et~al.}{2006}]{Nieten2006}
{Nieten} C.,  {Neininger} N.,  {Gu{\'e}lin} M.,  {Ungerechts} H.,  {Lucas} R.,
  {Berkhuijsen} E.~M.,  {Beck} R.,   {Wielebinski} R.,  2006, \mn@doi [\aap]
  {10.1051/0004-6361:20035672}, \href
  {https://ui.adsabs.harvard.edu/abs/2006A&A...453..459N} {453, 459}

\bibitem[\protect\citeauthoryear{{Obreschkow} \& {Rawlings}}{{Obreschkow} \&
  {Rawlings}}{2009}]{Obreschkow2009}
{Obreschkow} D.,  {Rawlings} S.,  2009, \mn@doi [\mnras]
  {10.1111/j.1365-2966.2009.14497.x}, \href
  {https://ui.adsabs.harvard.edu/abs/2009MNRAS.394.1857O} {394, 1857}

\bibitem[\protect\citeauthoryear{{Oh}, {Brook}, {Governato}, {Brinks}, {Mayer},
  {de Blok}, {Brooks}  \& {Walter}}{{Oh} et~al.}{2011}]{Oh2011}
{Oh} S.-H.,  {Brook} C.,  {Governato} F.,  {Brinks} E.,  {Mayer} L.,  {de Blok}
  W.~J.~G.,  {Brooks} A.,   {Walter} F.,  2011, \mn@doi [\aj]
  {10.1088/0004-6256/142/1/24}, \href
  {https://ui.adsabs.harvard.edu/abs/2011AJ....142...24O} {142, 24}

\bibitem[\protect\citeauthoryear{{Paladini}, {Montier}, {Giard}, {Bernard},
  {Dame}, {Ito}  \& {Macias-Perez}}{{Paladini} et~al.}{2007}]{Paladini2007}
{Paladini} R.,  {Montier} L.,  {Giard} M.,  {Bernard} J.~P.,  {Dame} T.~M.,
  {Ito} S.,   {Macias-Perez} J.~F.,  2007, \mn@doi [\aap]
  {10.1051/0004-6361:20065835}, \href
  {https://ui.adsabs.harvard.edu/abs/2007A&A...465..839P} {465, 839}

\bibitem[\protect\citeauthoryear{{Paturel}, {Theureau}, {Bottinelli},
  {Gouguenheim}, {Coudreau-Durand}, {Hallet}  \& {Petit}}{{Paturel}
  et~al.}{2003}]{Paturel2003}
{Paturel} G.,  {Theureau} G.,  {Bottinelli} L.,  {Gouguenheim} L.,
  {Coudreau-Durand} N.,  {Hallet} N.,   {Petit} C.,  2003, \mn@doi [\aap]
  {10.1051/0004-6361:20031412}, \href
  {https://ui.adsabs.harvard.edu/abs/2003A&A...412...57P} {412, 57}

\bibitem[\protect\citeauthoryear{{Peng}, {Wang}, {Liu}, {Tang}  \&
  {Wang}}{{Peng} et~al.}{2016}]{Peng2016}
{Peng} F.-K.,  {Wang} X.-Y.,  {Liu} R.-Y.,  {Tang} Q.-W.,   {Wang} J.-F.,
  2016, \mn@doi [\apjl] {10.3847/2041-8205/821/2/L20}, \href
  {https://ui.adsabs.harvard.edu/abs/2016ApJ...821L..20P} {821, L20}

\bibitem[\protect\citeauthoryear{{Peretti}, {Blasi}, {Aharonian}  \&
  {Morlino}}{{Peretti} et~al.}{2019}]{Peretti2019}
{Peretti} E.,  {Blasi} P.,  {Aharonian} F.,   {Morlino} G.,  2019, \mn@doi
  [\mnras] {10.1093/mnras/stz1161}, \href
  {https://ui.adsabs.harvard.edu/abs/2019MNRAS.487..168P} {487, 168}

\bibitem[\protect\citeauthoryear{{Persic}, {Rephaeli}  \& {Arieli}}{{Persic}
  et~al.}{2008}]{Persic2008}
{Persic} M.,  {Rephaeli} Y.,   {Arieli} Y.,  2008, \mn@doi [\aap]
  {10.1051/0004-6361:200809525}, \href
  {https://ui.adsabs.harvard.edu/abs/2008A&A...486..143P} {486, 143}

\bibitem[\protect\citeauthoryear{{Pfrommer}, {Pakmor}, {Simpson}  \&
  {Springel}}{{Pfrommer} et~al.}{2017}]{Pfrommer2017}
{Pfrommer} C.,  {Pakmor} R.,  {Simpson} C.~M.,   {Springel} V.,  2017, \mn@doi
  [\apjl] {10.3847/2041-8213/aa8bb1}, \href
  {https://ui.adsabs.harvard.edu/abs/2017ApJ...847L..13P} {847, L13}

\bibitem[\protect\citeauthoryear{{Pilyugin}, {Grebel}  \& {Kniazev}}{{Pilyugin}
  et~al.}{2014}]{Pilyugin2014}
{Pilyugin} L.~S.,  {Grebel} E.~K.,   {Kniazev} A.~Y.,  2014, \mn@doi [\aj]
  {10.1088/0004-6256/147/6/131}, \href
  {https://ui.adsabs.harvard.edu/abs/2014AJ....147..131P} {147, 131}

\bibitem[\protect\citeauthoryear{{Rekola}, {Richer}, {McCall}, {Valtonen},
  {Kotilainen}  \& {Flynn}}{{Rekola} et~al.}{2005}]{Rekola2005}
{Rekola} R.,  {Richer} M.~G.,  {McCall} M.~L.,  {Valtonen} M.~J.,  {Kotilainen}
  J.~K.,   {Flynn} C.,  2005, \mn@doi [\mnras]
  {10.1111/j.1365-2966.2005.09166.x}, \href
  {https://ui.adsabs.harvard.edu/abs/2005MNRAS.361..330R} {361, 330}

\bibitem[\protect\citeauthoryear{{R{\'e}my-Ruyer} et~al.,}{{R{\'e}my-Ruyer}
  et~al.}{2014}]{RemyRuyer2014}
{R{\'e}my-Ruyer} A.,  et~al., 2014, \mn@doi [\aap]
  {10.1051/0004-6361/201322803}, \href
  {https://ui.adsabs.harvard.edu/abs/2014A&A...563A..31R} {563, A31}

\bibitem[\protect\citeauthoryear{{R{\'e}my-Ruyer} et~al.,}{{R{\'e}my-Ruyer}
  et~al.}{2015}]{RemyRuyer2015}
{R{\'e}my-Ruyer} A.,  et~al., 2015, \mn@doi [\aap]
  {10.1051/0004-6361/201526067}, \href
  {https://ui.adsabs.harvard.edu/abs/2015A&A...582A.121R} {582, A121}

\bibitem[\protect\citeauthoryear{{S{\'a}nchez} et~al.,}{{S{\'a}nchez}
  et~al.}{2017}]{Sanchez2017}
{S{\'a}nchez} S.~F.,  et~al., 2017, \mn@doi [\mnras] {10.1093/mnras/stx808},
  \href {https://ui.adsabs.harvard.edu/abs/2017MNRAS.469.2121S} {469, 2121}

\bibitem[\protect\citeauthoryear{{Sanders}, {Mazzarella}, {Kim}, {Surace}  \&
  {Soifer}}{{Sanders} et~al.}{2003}]{Sanders2003}
{Sanders} D.~B.,  {Mazzarella} J.~M.,  {Kim} D.~C.,  {Surace} J.~A.,   {Soifer}
  B.~T.,  2003, \mn@doi [\aj] {10.1086/376841}, \href
  {https://ui.adsabs.harvard.edu/abs/2003AJ....126.1607S} {126, 1607}

\bibitem[\protect\citeauthoryear{{Sofue}, {Honma}  \& {Omodaka}}{{Sofue}
  et~al.}{2009}]{Sofue2009}
{Sofue} Y.,  {Honma} M.,   {Omodaka} T.,  2009, \mn@doi [\pasj]
  {10.1093/pasj/61.2.227}, \href
  {https://ui.adsabs.harvard.edu/abs/2009PASJ...61..227S} {61, 227}

\bibitem[\protect\citeauthoryear{{Springob}, {Haynes}, {Giovanelli}  \&
  {Kent}}{{Springob} et~al.}{2005}]{Springob2005}
{Springob} C.~M.,  {Haynes} M.~P.,  {Giovanelli} R.,   {Kent} B.~R.,  2005,
  \mn@doi [\apjs] {10.1086/431550}, \href
  {https://ui.adsabs.harvard.edu/abs/2005ApJS..160..149S} {160, 149}

\bibitem[\protect\citeauthoryear{{Stanimirovic}, {Staveley-Smith}, {Dickey},
  {Sault}  \& {Snowden}}{{Stanimirovic} et~al.}{1999}]{Stanimirovic1999}
{Stanimirovic} S.,  {Staveley-Smith} L.,  {Dickey} J.~M.,  {Sault} R.~J.,
  {Snowden} S.~L.,  1999, \mn@doi [\mnras] {10.1046/j.1365-8711.1999.02013.x},
  \href {https://ui.adsabs.harvard.edu/abs/1999MNRAS.302..417S} {302, 417}

\bibitem[\protect\citeauthoryear{{Staveley-Smith}, {Kim}, {Calabretta},
  {Haynes}  \& {Kesteven}}{{Staveley-Smith} et~al.}{2003}]{StaveleySmith2003}
{Staveley-Smith} L.,  {Kim} S.,  {Calabretta} M.~R.,  {Haynes} R.~F.,
  {Kesteven} M.~J.,  2003, \mn@doi [\mnras] {10.1046/j.1365-8711.2003.06146.x},
  \href {https://ui.adsabs.harvard.edu/abs/2003MNRAS.339...87S} {339, 87}

\bibitem[\protect\citeauthoryear{{Strickland}, {Heckman}, {Colbert}, {Hoopes}
  \& {Weaver}}{{Strickland} et~al.}{2004}]{Strickland2004}
{Strickland} D.~K.,  {Heckman} T.~M.,  {Colbert} E. J.~M.,  {Hoopes} C.~G.,
  {Weaver} K.~A.,  2004, \mn@doi [\apj] {10.1086/383136}, \href
  {https://ui.adsabs.harvard.edu/abs/2004ApJ...606..829S} {606, 829}

\bibitem[\protect\citeauthoryear{{Strong}, {Porter}, {Digel},
  {J{\'o}hannesson}, {Martin}, {Moskalenko}, {Murphy}  \& {Orlando}}{{Strong}
  et~al.}{2010}]{strong2010}
{Strong} A.~W.,  {Porter} T.~A.,  {Digel} S.~W.,  {J{\'o}hannesson} G.,
  {Martin} P.,  {Moskalenko} I.~V.,  {Murphy} E.~J.,   {Orlando} E.,  2010,
  \mn@doi [\apjl] {10.1088/2041-8205/722/1/L58}, \href
  {https://ui.adsabs.harvard.edu/abs/2010ApJ...722L..58S} {722, L58}

\bibitem[\protect\citeauthoryear{{Sudoh}, {Totani}  \& {Kawanaka}}{{Sudoh}
  et~al.}{2018}]{Sudoh2018}
{Sudoh} T.,  {Totani} T.,   {Kawanaka} N.,  2018, \mn@doi [\pasj]
  {10.1093/pasj/psy039}, \href
  {https://ui.adsabs.harvard.edu/abs/2018PASJ...70...49S} {70, 49}

\bibitem[\protect\citeauthoryear{{Sudoh}, {Linden}  \& {Beacom}}{{Sudoh}
  et~al.}{2019}]{2019PhRvD.100d3016S}
{Sudoh} T.,  {Linden} T.,   {Beacom} J.~F.,  2019, \mn@doi [\prd]
  {10.1103/PhysRevD.100.043016}, \href
  {https://ui.adsabs.harvard.edu/abs/2019PhRvD.100d3016S} {100, 043016}

\bibitem[\protect\citeauthoryear{{Tang}, {Wang}  \& {Tam}}{{Tang}
  et~al.}{2014}]{Tang2014}
{Tang} Q.-W.,  {Wang} X.-Y.,   {Tam} P.-H.~T.,  2014, \mn@doi [\apj]
  {10.1088/0004-637X/794/1/26}, \href
  {https://ui.adsabs.harvard.edu/abs/2014ApJ...794...26T} {794, 26}

\bibitem[\protect\citeauthoryear{{Torres}}{{Torres}}{2004}]{Torres2004}
{Torres} D.~F.,  2004, \mn@doi [\apj] {10.1086/425415}, \href
  {https://ui.adsabs.harvard.edu/abs/2004ApJ...617..966T} {617, 966}

\bibitem[\protect\citeauthoryear{{VERITAS Collaboration} et~al.,}{{VERITAS
  Collaboration} et~al.}{2009}]{VERITASCollaboration2009}
{VERITAS Collaboration} et~al., 2009, \mn@doi [\nat] {10.1038/nature08557},
  \href {https://ui.adsabs.harvard.edu/abs/2009Natur.462..770V} {462, 770}

\bibitem[\protect\citeauthoryear{{Wang} \& {Fields}}{{Wang} \&
  {Fields}}{2018}]{Wang2018}
{Wang} X.,  {Fields} B.~D.,  2018, \mn@doi [\mnras] {10.1093/mnras/stx2917},
  \href {https://ui.adsabs.harvard.edu/abs/2018MNRAS.474.4073W} {474, 4073}

\bibitem[\protect\citeauthoryear{{Wei{\ss}}, {Walter}  \&
  {Scoville}}{{Wei{\ss}} et~al.}{2005}]{Weis2005}
{Wei{\ss}} A.,  {Walter} F.,   {Scoville} N.~Z.,  2005, \mn@doi [\aap]
  {10.1051/0004-6361:20052667}, \href
  {https://ui.adsabs.harvard.edu/abs/2005A&A...438..533W} {438, 533}

\bibitem[\protect\citeauthoryear{{Xi}, {Liu}, {Wang}, {Yang}, {Yuan}  \&
  {Zhang}}{{Xi} et~al.}{2020a}]{Xi2020b}
{Xi} S.-Q.,  {Liu} R.-Y.,  {Wang} X.-Y.,  {Yang} R.-Z.,  {Yuan} Q.,   {Zhang}
  B.,  2020a, \mn@doi [\apjl] {10.3847/2041-8213/ab982c}, \href
  {https://ui.adsabs.harvard.edu/abs/2020ApJ...896L..33X} {896, L33}

\bibitem[\protect\citeauthoryear{{Xi}, {Zhang}, {Liu}  \& {Wang}}{{Xi}
  et~al.}{2020b}]{Xi2020}
{Xi} S.-Q.,  {Zhang} H.-M.,  {Liu} R.-Y.,   {Wang} X.-Y.,  2020b, \mn@doi
  [\apj] {10.3847/1538-4357/aba043}, \href
  {https://ui.adsabs.harvard.edu/abs/2020ApJ...901..158X} {901, 158}

\bibitem[\protect\citeauthoryear{{Yoast-Hull}, {Everett}, {Gallagher}  \&
  {Zweibel}}{{Yoast-Hull} et~al.}{2013}]{YoastHull2013}
{Yoast-Hull} T.~M.,  {Everett} J.~E.,  {Gallagher} J.~S. I.,   {Zweibel} E.~G.,
   2013, \mn@doi [\apj] {10.1088/0004-637X/768/1/53}, \href
  {https://ui.adsabs.harvard.edu/abs/2013ApJ...768...53Y} {768, 53}

\bibitem[\protect\citeauthoryear{{Yoast-Hull}, {Gallagher}, {Zweibel}  \&
  {Everett}}{{Yoast-Hull} et~al.}{2014}]{YoastHull2014}
{Yoast-Hull} T.~M.,  {Gallagher} J.~S. I.,  {Zweibel} E.~G.,   {Everett} J.~E.,
   2014, \mn@doi [\apj] {10.1088/0004-637X/780/2/137}, \href
  {https://ui.adsabs.harvard.edu/abs/2014ApJ...780..137Y} {780, 137}

\bibitem[\protect\citeauthoryear{{Yoast-Hull}, {Gallagher}, {Aalto}  \&
  {Varenius}}{{Yoast-Hull} et~al.}{2017}]{YoastHull2017}
{Yoast-Hull} T.~M.,  {Gallagher} John~S. I.,  {Aalto} S.,   {Varenius} E.,
  2017, \mn@doi [\mnras] {10.1093/mnrasl/slx054}, \href
  {https://ui.adsabs.harvard.edu/abs/2017MNRAS.469L..89Y} {469, L89}

\bibitem[\protect\citeauthoryear{{Young}, {Claussen}, {Kleinmann}, {Rubin}  \&
  {Scoville}}{{Young} et~al.}{1988}]{Young1988}
{Young} J.~S.,  {Claussen} M.~J.,  {Kleinmann} S.~G.,  {Rubin} V.~C.,
  {Scoville} N.,  1988, \mn@doi [\apjl] {10.1086/185240}, \href
  {https://ui.adsabs.harvard.edu/abs/1988ApJ...331L..81Y} {331, L81}

\bibitem[\protect\citeauthoryear{{Young}, {Xie}, {Kenney}  \& {Rice}}{{Young}
  et~al.}{1989}]{Young1989}
{Young} J.~S.,  {Xie} S.,  {Kenney} J. D.~P.,   {Rice} W.~L.,  1989, \mn@doi
  [\apjs] {10.1086/191355}, \href
  {https://ui.adsabs.harvard.edu/abs/1989ApJS...70..699Y} {70, 699}

\bibitem[\protect\citeauthoryear{{Young} et~al.,}{{Young}
  et~al.}{1995}]{Young1995}
{Young} J.~S.,  et~al., 1995, \mn@doi [\apjs] {10.1086/192159}, \href
  {https://ui.adsabs.harvard.edu/abs/1995ApJS...98..219Y} {98, 219}

\bibitem[\protect\citeauthoryear{{van der Marel}}{{van der
  Marel}}{2006}]{vanderMarel2006}
{van der Marel} R.~P.,  2006, in {Livio} M.,  {Brown} T.~M.,  eds, The Local
  Group as an Astrophysical Laboratory. pp 47--71 (\mn@eprint {arXiv}
  {astro-ph/0404192})

\makeatother
\end{thebibliography}


\appendix
\section{Model predictions for galaxies removed from our sample}
\label{section:appendix}

In this section, we show the predicted spectra of galaxies removed from our original sample
for some reasons (Section \ref{section:sample}).
Because of the lack of information necessary to make our model predictions,
not all of the removed galaxies are considered here.

M31 was removed from our sample due to the discrepancy between 
the regions of gas and gamma-ray emission. However, 
the predicted flux and spectrum are in good agreement with observations. 
This may indicate that the ring-like gas distribution \citep{Ackermann2017},
which is not taken into account in our model, does not seriously 
affect gamma-ray emission properties. 

NGC 1068 and Arp 220, were removed from our sample due to the possibilities of AGN activities. 
Predicted spectra by our model are lower than observed fluxes, which is consistent 
if the AGN activities are dominant in gamma-ray emission in these galaxies. 
Note that the predicted flux would become even smaller for Arp 220 
if we consider gamma-ray absorption by pair-production in the galaxy.

\begin{figure*}
\begin{minipage}{0.33\hsize}
\begin{center}
\includegraphics[width=\linewidth]{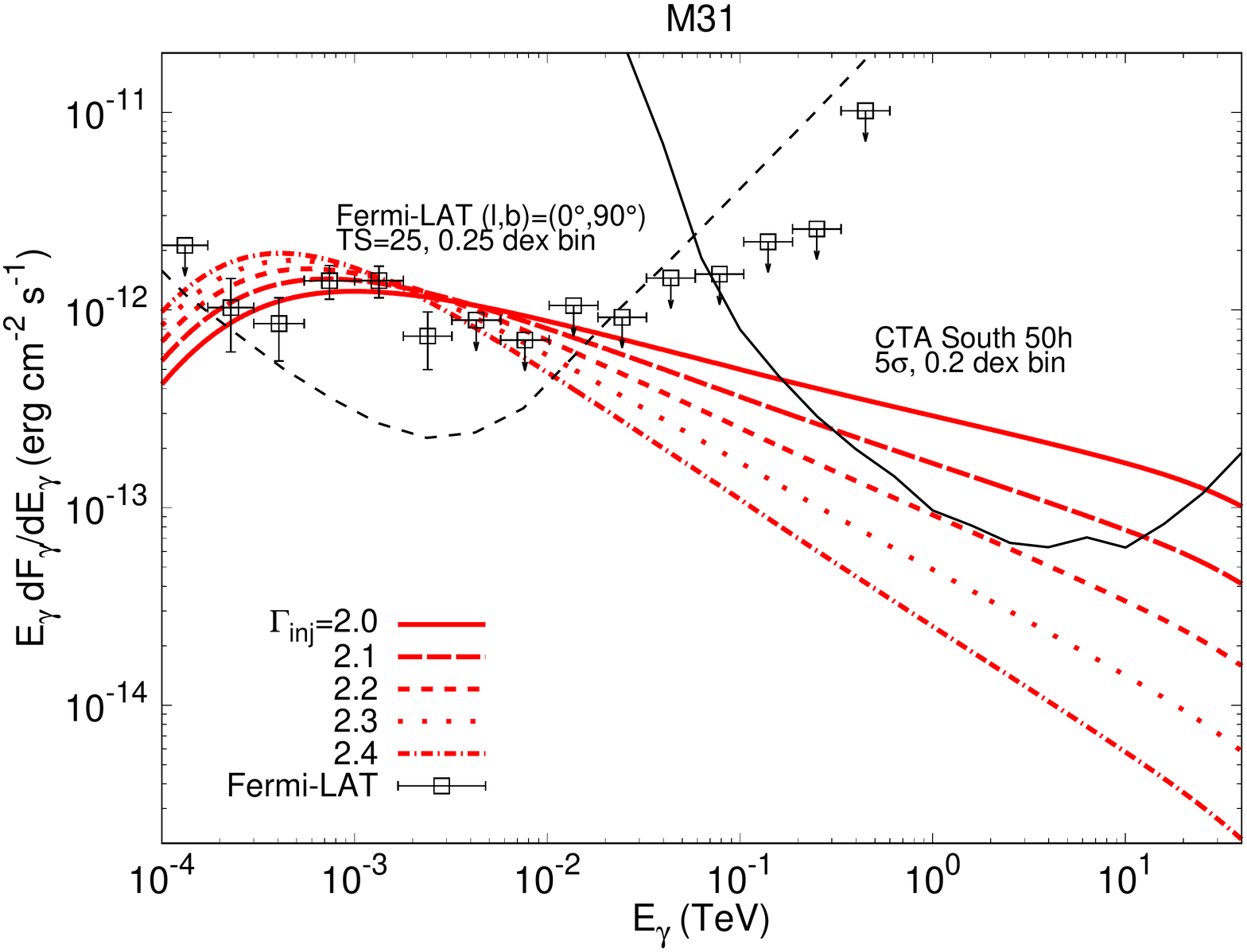}
\end{center}
\end{minipage}
\begin{minipage}{0.33\hsize}
\begin{center}
\includegraphics[width=\linewidth]{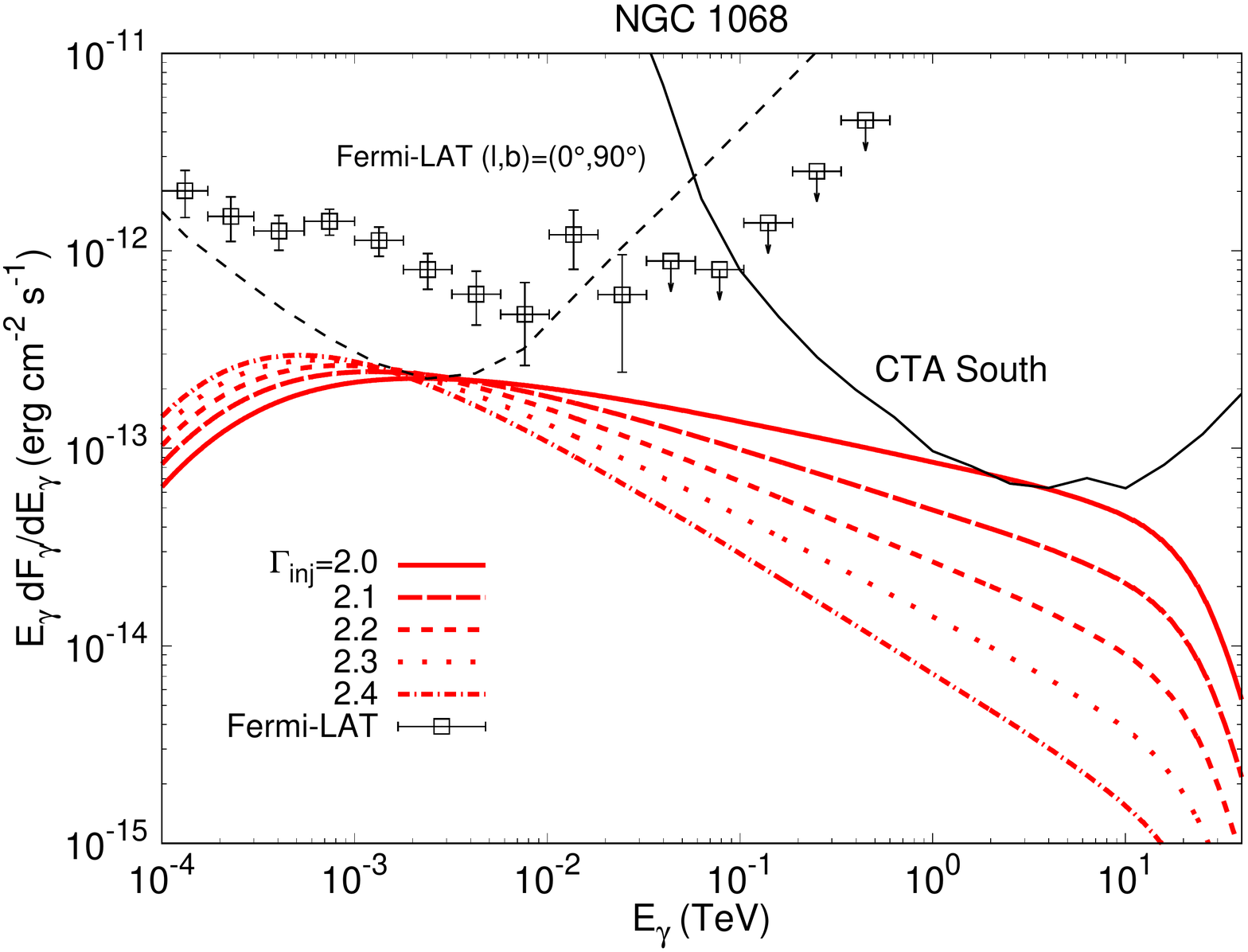}
\end{center}
\end{minipage}
\begin{minipage}{0.33\hsize}
\begin{center}
\includegraphics[width=\linewidth]{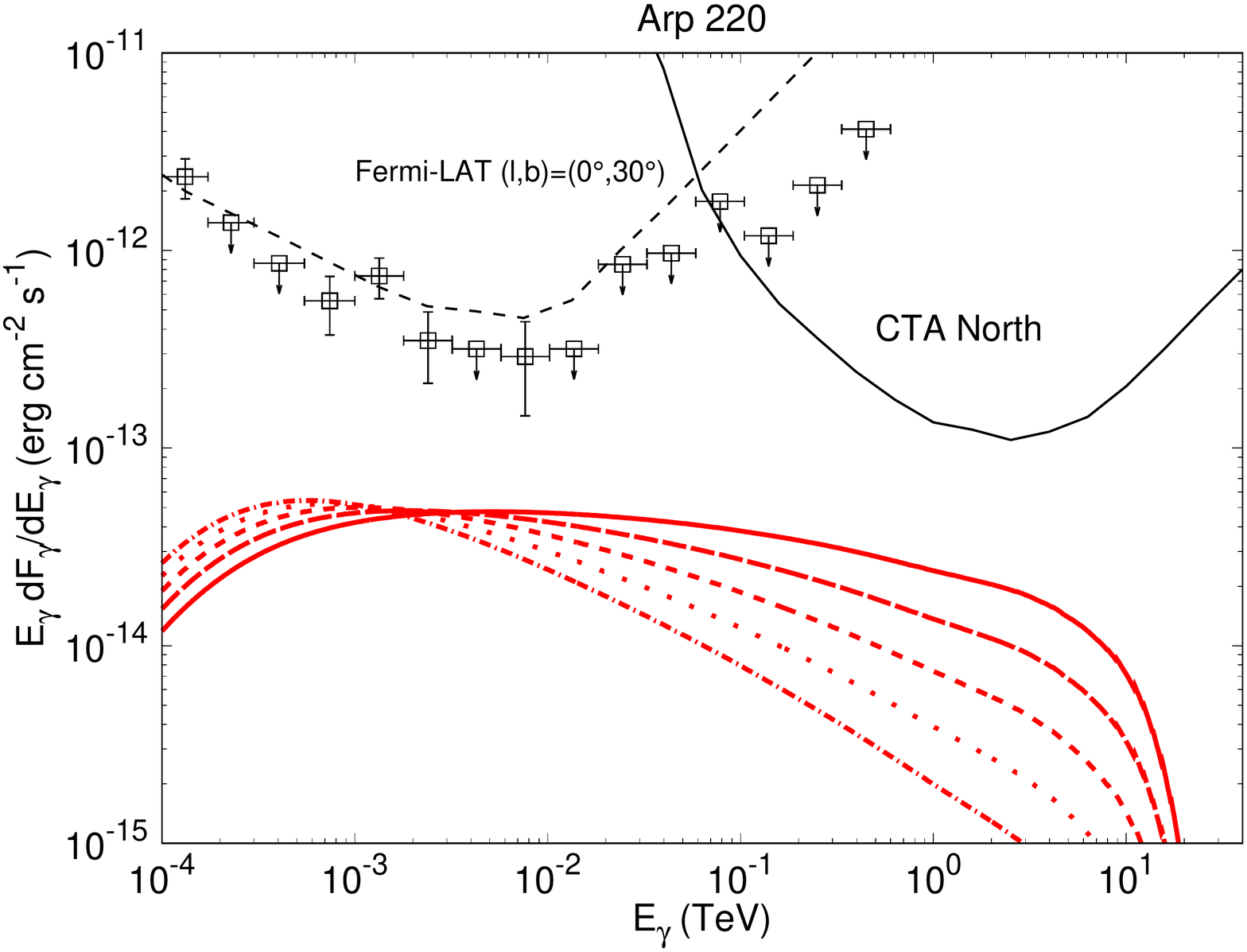}
\end{center}
\end{minipage}
\caption{
Predicted gamma-ray spectra of M31, NGC 1068, Arp 220
with the observed spectra \citep{Ajello2020}.
}
\label{fig:extended}
\end{figure*}

\begin{table*}
\begin{minipage}{\textwidth}
\renewcommand\thefootnote{\fnsymbol{footnote}}
\begin{center}
\caption{Physical parameters of M31, NGC 1068, and Arp 220.}
\label{tab:appendix1}
\begin{tabular}{ccccccccccc}\hline
name &
$D\medspace (\mathrm{Mpc})$&ref&
$\psi (\mathrm{M_\odot\,yr^{-1}})$&ref&
$M_\mathrm{gas}\medspace (10^9\medspace \mathrm{M_\odot})$&ref&
$M_\mathrm{star}\medspace (10^9\medspace \mathrm{M_\odot})$&ref&
$R_\mathrm{eff}\medspace (\mathrm{kpc})$&ref\\
\hline\hline
M31 & 0.79 & (1) & 0.53 & (1),(2) & 7.6 & (3),(4),(5) & 86 & (6) & 1.8 & (7)\\
NGC 1068 & 14 & (2) & 17 & (2),(8) & 11 & (5),(9),(10) & 74 & (11) & 1.5 & (7)\\
Arp 220 & 80 & (3) & 98 & (2),(12) & 51 & (10),(13),(14) & 61 & (15) & 3.5 & (7)\\

\hline
\end{tabular}
\footnotetext{
References: (1) \citet{Kennicutt2008}, (2) \citet{Sanders2003}, (3) \citet{Braun2009}, (4) \citet{Nieten2006}, (5) \citet{Pilyugin2014}, 
(6) \citet{Montalto2009}, (7) \citet{Jarrett2003}, (8) \citet{Adams1975}, (9) \citet{Haynes2018}, (10) \citet{Young1995},
(11) \citet{Howell2007}, (12) \citet{Armus1990}, (13) \citet{Paturel2003}, (14) \citet{Sanchez2017}, (15) \citet{Brown2014},
}
\end{center}
\end{minipage}
\end{table*}

\begin{table*}
\begin{center}
  \caption{List of gamma-ray fluxes ($E_\gamma \, dF_\gamma/dE_\gamma$ at
    1 TeV) predicted by our model, in units of
    $\mathrm{erg\,s^{-1}\,cm^{-2}}$ assuming
    $\Gamma_{\rm inj} = 2.2$.}
\label{tab:appendix2}
\begin{tabular}{|l|ccccc|}
\hline
name&$\Gamma_\mathrm{inj}=2.0$&$\Gamma_\mathrm{inj}=2.1$&$\Gamma_\mathrm{inj}=2.2$&$\Gamma_\mathrm{inj}=2.3$&$\Gamma_\mathrm{inj}=2.4$\\
\hline\hline

M31 & $2.9\times10^{-13}$ & $1.7\times10^{-13}$ & $9.2\times10^{-14}$ & $4.9\times10^{-14}$ & $2.5\times10^{-14}$\\
NGC 1068 & $8.5\times10^{-14}$ & $4.9\times10^{-14}$ & $2.7\times10^{-14}$ & $1.4\times10^{-14}$ & $7.2\times10^{-15}$\\
Arp 220 & $2.4\times10^{-14}$ & $1.4\times10^{-14}$ & $7.4\times10^{-15}$ & $3.9\times10^{-15}$ & $2.0\times10^{-15}$\\

\hline
\end{tabular}
\end{center}
\end{table*}


\bsp	
\label{lastpage}
\end{document}